\documentclass[sigconf]{acmart}
\acmConference[MSR 2022]{MSR '22: Proceedings of the 19th International Conference on Mining Software Repositories}{May 23–24, 2022}{Pittsburgh, PA, USA}
\AtBeginDocument{%
  \providecommand\BibTeX{{%
    \normalfont B\kern-0.5em{\scshape i\kern-0.25em b}\kern-0.8em\TeX}}}

\setcopyright{acmcopyright}
\copyrightyear{2022} 
\acmYear{2022} 
\setcopyright{acmlicensed}\acmConference[MSR '22]{19th International Conference on Mining Software Repositories}{May 23--24, 2022}{Pittsburgh, PA, USA}
\acmBooktitle{19th International Conference on Mining Software Repositories (MSR '22), May 23--24, 2022, Pittsburgh, PA, USA}
\acmPrice{15.00}
\acmDOI{10.1145/3524842.3527947}
\acmISBN{978-1-4503-9303-4/22/05}

\acmConference[MSR 2022]{MSR '22: Proceedings of the 19th International Conference on Mining Software Repositories}{May 23–24, 2022}{Pittsburgh, PA, USA}

\usepackage{xspace}
\usepackage{bm}
\usepackage{adjustbox}
\usepackage{tabularx}
\usepackage{color}
\usepackage{algorithm}
\usepackage{algpseudocode}
\usepackage{caption}
\usepackage{subcaption}
\usepackage{listings}
\usepackage{nccmath}
\usepackage[normalem]{ulem}

\begin{document}

\title{Senatus - A Fast and Accurate Code-to-Code Recommendation Engine}

%

%
\author{Fran Silavong}
\affiliation{%
  \institution{CTO, JPMorgan Chase}
  \city{London}
  \country{United Kingdom}}
\email{fran.silavong@jpmchase.com}

\author{Sean Moran}
\affiliation{%
  \institution{CTO, JPMorgan Chase}
  \city{London}
  \country{United Kingdom}}
\email{sean.j.moran@jpmchase.com}

\author{Antonios Georgiadis}
\affiliation{%
  \institution{CTO, JPMorgan Chase}
  \city{London}
  \country{United Kingdom}}
\email{antonios.georgiadis@jpmchase.com}

\author{Rohan Saphal}
\affiliation{%
  \institution{CTO, JPMorgan Chase}
  \city{Glasgow}
  \country{United Kingdom}}
\email{rohan.saphal@jpmchase.com}

\author{Robert Otter}
\affiliation{%
  \institution{CTO, JPMorgan Chase}
  \city{London}
  \country{United Kingdom}}
\email{robert.otter@jpmchase.com}

\makeatletter
\DeclareRobustCommand\onedot{\futurelet\@let@token\@onedot} 
\def\@onedot{\ifx\@let@token.\else.\null\fi\xspace}

\def\eg{\emph{e.g}\onedot} \def\Eg{\emph{E.g}\onedot}
\def\ie{\emph{i.e}\onedot} \def\Ie{\emph{I.e}\onedot}
\def\cf{\emph{cf}\onedot} \def\Cf{\emph{Cf}\onedot}
\def\etc{\emph{etc}\onedot} \def\vs{\emph{vs}\onedot}
\def\wrt{w.r.t\onedot} \def\dof{d.o.f\onedot}
\def\etal{\emph{et al}\onedot}
\makeatother

\lstdefinestyle{codestyle}{ 
    keywordstyle=\color{blue},
    breaklines=True,
    basicstyle=\scriptsize,
}

\lstset{style=codestyle}

\definecolor{seancolor}{rgb}{0.1,0.1,0.8}
\newcommand{\sean}[1]{{\textcolor{seancolor}{[SJM #1]}}}
\newcommand\sjm[1] {\marco{#1}}

\definecolor{francolor}{rgb}{0.8,0.1,0.1}
\newcommand{\fran}[1]{{\textcolor{francolor}{[FS #1]}}}
\newcommand\fs[1] {\fran{#1}}

\definecolor{tonycolor}{rgb}{0.8,0.8,0.1}
\newcommand{\tony}[1]{{\textcolor{tonycolor}{[AG #1]}}}
\newcommand\ag[1] {\tony{#1}}

\begin{abstract}
  Machine learning on source code (MLOnCode) is a popular research field that has been driven by the availability of large-scale
  code repositories and the development of powerful probabilistic and deep learning models for mining source code. Code-to-code recommendation is a task in MLOnCode that aims to recommend relevant, diverse and concise code snippets that usefully extend the code currently being written by a developer in their development environment (IDE). Code-to-code recommendation engines hold the promise of increasing developer productivity by reducing context switching from the IDE and increasing code-reuse. Existing code-to-code recommendation engines do not scale gracefully to large codebases, exhibiting a linear growth in query time as the code repository increases in size. In addition, existing code-to-code recommendation engines fail to account for the global statistics of code repositories in the ranking function, such as the distribution of code snippet lengths, leading to sub-optimal retrieval results. We address both of these weaknesses with \emph{Senatus}, a new code-to-code recommendation engine. At the core of Senatus is \emph{De-Skew} LSH a new locality sensitive hashing (LSH) algorithm that indexes the data for fast (sub-linear time) retrieval while also counteracting the skewness in the snippet length distribution using novel abstract syntax tree-based feature scoring and selection algorithms. We evaluate Senatus and find the recommendations to be of higher quality than competing baselines, while achieving faster search. For example on the CodeSearchNet dataset Senatus improves performance by 31.21\% F1 and  147.9\emph{x} faster query time compared to Facebook Aroma. Senatus also outperforms standard MinHash LSH by 29.2\% F1 and 51.02\emph{x} faster query time. 
\end{abstract}





\maketitle

\section{Introduction}

High quality and reliable software is a must for the smooth running of a multitude of systems that drive the modern world, from the financial system, online commerce to landing the latest exploratory rover on Mars. Machine learning for mining code repositories (MLOnCode) has gained traction  in the research community with the emergence of large-scale repositories of opensource code (\eg GitHub, CodeNet~\cite{puri2021codenet}), with associated metadata (\eg comments, documentation, commit history, resource usage) and the parallel advancements in the development of probabilistic and deep learning models for pattern recognition in source code~\cite{allamanis2018survey}. There is scope for MLOnCode to facilitate the familiar software development experience by augmenting common tasks with machine learning to enhance developer productivity, efficiency and improve code reliability and consistency. For example, in the build stage of the Software Development Lifecyle (SDLC), code-to-code recommendation engines~\cite{luan2019,li2018,bhoopchand2016learning,Cambronero19} suggest contextually relevant code directly within the developer's IDE, reducing the need to context switch and leave the IDE to search the web. The promise of code-to-code recommendation engines is built on the finding that most code is similar to code that is already written, for example in one study up to 40\% of the code repositories were found to have noticeable overlap~\cite{luan2019}. In another example, the testing stage of the SDLC is aided by automatic unit test generation tools~\cite{PachecoLEB2007,tufano2020unit} that are able to automatically generate a bank of unit tests and program repair and bug detection innovations~\cite{le2019,Pradel2018,Le19,Zhou17}.

In this paper we focus on the build and test phase of the SDLC lifecycle, and the specific task of code-to-code recommendation. We follow the definition given by Luan~\etal~\cite{luan2019} in which the input to the search engine is a code snippet and the output is a ranked list of \emph{concise}, \emph{diverse} and \emph{relevant} code snippets that \emph{usefully extend the query code snippet with additional functionality}\footnote{A code snippet is defined as either a complete method or function of arbitrary length when referring to the corpus, or a piece of complete or semi-complete code of arbitrary length when referring to the query.}. This definition is a twist on the familiar task of code search which typically does not require the snippets in the ranked list to extend the query snippet or to be concise and diverse.
Code search is a well researched field with a multitude of prior work in the area, roughly broken down into code clone detectors and code-search tools~\cite{Cordy11,Jiang07,Kamiya02,Sajani16,luan2019,li2018,bhoopchand2016learning,Cambronero19,Balachandran15,Shuhan20,deRezende20,Wei20,Krugler2013,Kim18}. Code-to-code recommendation can be approached using existing systems for code search and clone detection although both types of systems have their disadvantages for code-to-code recommendation as highlighted by Luan~\etal~\cite{luan2019}. For example, code clone detectors~\cite{Cordy11,Jiang07,Kamiya02,Sajani16} retrieve mostly near duplicate or duplicate snippets which are not useful in code-to-code recommendation and code search tools~\cite{Krugler2013,Kim18,Balachandran15} make no attempt at ensuring that the ranked snippets usefully extend the query. In evaluating the aforementioned solutions for code-to-code recommendation, search and clone detection we generally find that there is a gap in scaling these systems to massive codebases while simultaneously maintaining or even improving the retrieval quality of the system. 
In addition to scalability concerns, we also find limitations in existing code search tools in terms of retrieval effectiveness. Code-to-code recommendation is a challenging task for contemporary Information Retrieval (IR) methods; code is designed to be understood by both humans and by machine and this requirement leads to the similarities between code and natural language~\cite{allamanis2018survey} that can be exploited by traditional text-based search engines. Nevertheless there are many key differences between natural language and code which the search engine should ideally take into consideration when ranking. For example code has well-defined structure, a formal syntax and semantics and is designed from the start to be executed~\cite{allamanis2018survey}. Treating code as a bag-of-words and applying traditional search techniques such as BM25 and TF-IDF lead to sub-optimal performance compared to search that encapsulates the logical structure of the code in the featurisation~\cite{luan2019}. In addition to code properties such as the logical structure, in this paper we also argue that is important for a retrieval engine to consider the global statistics of the data in a code repository. More specifically, we provide evidence of a \emph{power-law distribution in code snippet lengths} in typical code repositories, in which there is a large number of short code snippets with a long tail of larger code snippets. Existing code search tools do not take this phenomenon into consideration when ranking, leading to a degradation in retrieval effectiveness.

In this work we propose \emph{Senatus}, a new code-to-code recommendation engine that addresses both of these issues. Senatus exhibits a sub-linear growth in query-time with respect to the code repository size while addressing retrieval quality degradation by novel feature scoring and selection techniques that counteracts the data skewness inherent in typical code repositories with respect to snippet length. Senatus works by indexing the code repository using Minhash-LSH. Minwise Hashing (Minhash)~\cite{Broder98,Broder97,shrivastava2015asymmetric, Ping11} is a well-known method for maintaining the Jaccard similiarity between sets by re-representing those sets as typically much lower-dimensional similarity preserving feature vectors. Locality sensitive hashing (LSH)~\cite{Andoni08,Indyk98} as applied to those Minhash feature vectors can provide an effective technique for indexing the data so that, at query time, only buckets to which the query hashes need to be checked for similar items. Intuitively, items colliding in the same buckets should be highly likely to have a high Jaccard similarity thereby circumventing a search across all items in the databasse. Minhash-LSH has been applied to a large range of applications from web search~\cite{Zhu16}, earthquake detection~\cite{Rong18},  graph sampling~\cite{Cormode05} to scalable online collaborative filtering for news personalisation~\cite{Das07} and efficient clustering of massive collections of genomic sequences~\cite{Ondov16}. Different to these existing approaches, we explore the application of Minhash-LSH to the task of code-to-code recommendation over large-scale code repositories. We find that Minhash-LSH has poor retrieval quality  out-of-the-box when applied to the code-to-code recommendation task and we identify the aforementioned skewness in code snippet lengths to be the primary issue. Senatus counteracts the data skewness by applying novel feature scoring and selection methods that operate directly on the abstract syntax tree (AST) structural representation of code, leading to both faster query time and superior retrieval effectiveness.

Our contributions in this paper are three-fold: 

\begin{itemize}
\item \textbf{Scalable Code-to-Code Recommendation:} We introduce \textit{DeSkew-LSH} a new method for approximate nearest neighbour search that accelerates the lookup time for code-to-code-recommendation. De-Skew LSH achieves a $147.9x$ improvement in query time without the requirement for any specialized hardware, and $+31.21\%$ improvement in retrieval effectiveness versus Facebook's state-of-the-art Aroma code recommendation engine. The code will be opensourced.
\item \textbf{Data Skewness in Code Repositories:} We study the \emph{distribution of code snippet lengths} in two public source code repositories, CodeSearchNet~\cite{husain2020codesearchnet} and the Neural Code Search evaluation dataset~\cite{li2019neural} and find that the snippet lengths are heavily skewed, following a power-law distribution, with the vast majority of the snippets being short in length, and a long tail of longer snippets. We argue that code-to-code recommendation engines, to return concise and useful snippets, should implement techniques to counteract the bias caused by this skewness.
\item \textbf{Feature Scoring and Selection:} Based on the skewness findings, we propose two novel feature scoring methods, Normalized Sub-Path Frequency (NSPF) and Inverse Leaves Frequency (ILF) that score parse tree features based on the tree structure in an Abstract Syntax Tree (AST) parsing of a code snippet. We argue that these tree structure sensitive feature scoring methods are more suitable for tackling skewness for the purposes of code-to-code recommendation. Based on the scores produced by these functions we propose two feature selection techniques: Top-F and Mid-C percentile for selecting a compact, discriminative set of code features based on the scores.
\end{itemize}


\section{Preliminaries} \label{sec:definitions}

In this section we describe how to represent code as a set and manipulate those sets to achieve fast similarity search. We  describe a previously known method, Simplified Parse Trees (SPTs)~\cite{luan2019}, for representing code snippets as \emph{sets}. This set representation of code is a pre-requisite for application of MinHash-LSH. Minwise hashing~\cite{Broder97} and locality sensitive hashing (LSH)~\cite{Indyk98} are fundamental foundations of Senatus that enable fast search over set-based data. The reference material~\cite{rajaraman2012mining} provides further background on Minhash-LSH. 

\subsection{Simplified Parse Tree (SPT) Code Representation}~\label{sec:spt}

We first describe how we featurise code snippets into sets so that they can be compared for similarity. We follow the Simplified Parse Tree (SPT) featurisation technique proposed by Aroma~\cite{luan2019}, although we note that many other possibilities exist for creating code vectors~\eg\cite{Alon19,zuegner_code_transformer_2021}. The Aroma code search engine transforms each code snippet into a SPT for the purposes of code similarity computation. The SPT is a representation of the code snippet that summarizes the salient structural information relating to the snippet. The SPT is designed to remove irrelevant lower level syntactical information from the snippet as well as language-specific tokens and thus better captures the program structure rather than program syntax, as recommended by Aroma. In our work, \emph{structural features}, namely token, parent, sibling, variable usage features, extracted from a traversal of the SPT are used to represent the code snippets for the purposes of similarity computation.

Code snippets (\eg queries, methods or classes) are converted to their corresponding SPT and the featurisation algorithm traverses through the tree to extract four types of features: (1) Token Features, (2) Parent Features (3) Sibling Features, and (4) Variable Usage Features. More specifically, the source code first gets converted to an AST using ANTLR with the appropriate Lexer and Parser grammars for the language of choice. We implemented our solution in Java as it is a popular, platform-independent language, however the featurisation would be compatible with other languages so long as an AST can be extracted. The ASTs are then simplified to SPTs by replacing local variables with the \#VAR token, unless they are method names or global variables. In addition:
\vspace{-0.03in}

\begin{itemize}
  \item SPTs are represented as a list (or array) of subtrees and leaf nodes.
  \item Subtrees are represented by one or more "$\#$".
  \item Parent relationships are represented using "$>$", and are not limited only to direct parent.
  \item Sibling relationships are represented using "$>>$".
  \item Variable re-use is represented using "$>>>$".
  \item Numerical indices are used to differentiate relationships of the same type (\eg parent from parent, sibling from sibling).
\end{itemize}

The parent and sibling features help to capture the local relations between different pairs of nodes and  provide a weak structural signal. If only token features were used it would be hard to distinguish if the feature set was extracted from a tree representation or from a text sequence. 

The token, parent, sibling and variable features form the dimensions in our vector space. With the dimensions defined, we create binary feature vectors for each snippet by recording the absence or presence of a feature in the snippet's SPT representation. The resulting binary vectors are used in similarity comparisons in Senatus to perform ranked retrieval of code as we describe in Section~\ref{sec:deskew}.

\begin{figure*}
  \includegraphics[width=\textwidth]{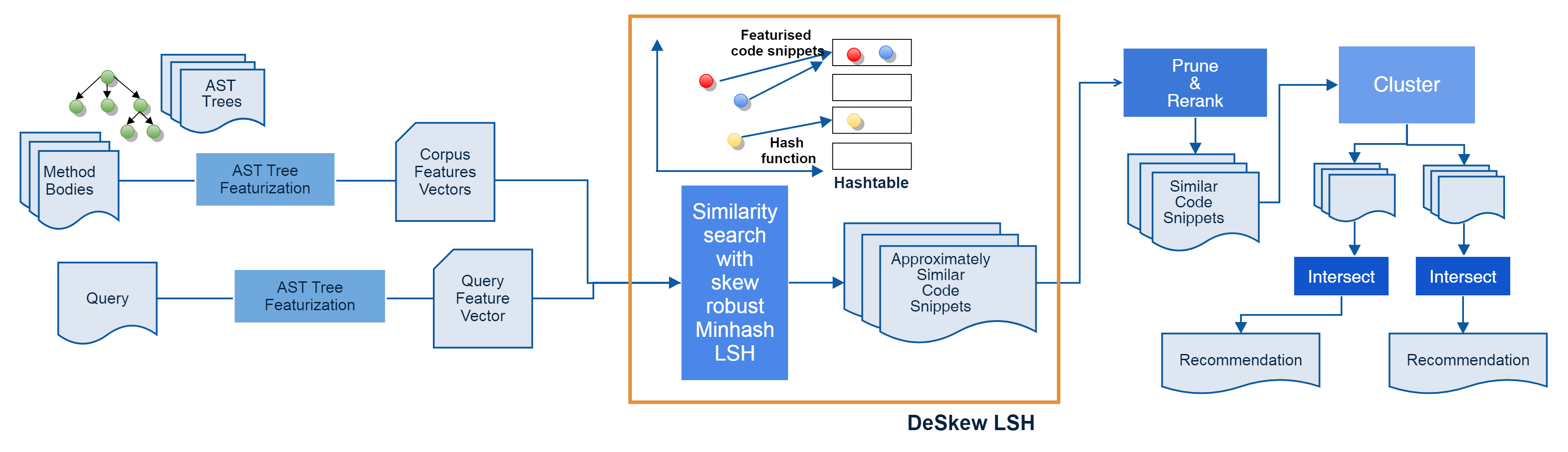}
  \caption{Overview of \emph{Senatus}.  Method bodies are featurized by traversing their abstract syntax tree (AST). Our proposed feature selection techniques are applied to the resulting features, which are then minhashed to produce signatures. LSH is used to bucket the signatures, with colliding signatures returned as approximately similar snippets that are passed onto the downstream tasks to generate recommendations for the end-user. The orange box indicates the De-Skew LSH stage of \emph{Senatus}.}
  \Description{Senatus}
  \label{fig:senatus}
\end{figure*}

\subsection{Minwise Hashing (Minhash)}\label{sec:minhash}

Minwise hashing (Minhash) is an algorithm originally proposed by Broder~\etal~\cite{Broder97} that transforms sets into compact signatures that can be used to estimate the Jaccard similarity between the sets. In Senatus, code snippets are converted into sets by the procedure outlined in Section~\ref{sec:spt}. Jaccard similarity measures the overlap between two sets and has been used with success to compare web documents~\cite{Broder97} and genomic analysis~\cite{Rangwala2013MCMinHMC}, amongst many other applications. The Minhash signatures can be much more compact than the underlying sets, saving memory. In this paper we explore the application of Jaccard similarity for the task of code-to-code recommendation.

The objective of minhash is to transform a set into a signature such that we can estimate the Jaccard similarity of sets just by using their signatures, the original sets are no longer needed. Furthermore, as the length of the signature increases, the estimate of the Jaccard similarity should become more accurate. To achieve these properties, minhash uses a set of $K$ random hash functions $h_{i}$ that takes as input an element of the set and returns an integer hash value. To obtain the minhash for the set for hash function $h_{i}$, the hash function is applied to each item in the set and the minimum hash value is returned, which we denote as $h^{min}_{i}(s)$. This is the first component of the minhash signature for the set. Additional components can be generated by using more random hash functions. The usefulness of minhashing is the connection between minhashing and the Jaccard similarity of sets. The probability that the minimum values returned from the application of a hash function to two sets $s_{1}, s_{2}$ or $Pr(h^{min}_{i}(s_{1})=h^{min}_{i}(s_{2}))$ is equal to the Jaccard similarity between the sets~\cite{Broder98}. A signature for a set $s$ is a vector of the minhash values resulting from $K$ independent hash functions: $[h^{min}_{1}(s), h^{min}_{2}(s),\ldots,h^{min}_{K}(s)]$. To estimate the Jaccard similarity between two sets we can compare their minhash values, enumerating the number of matching values, and dividing by the total length of a signature. Minwise hashing provides the signatures for our sets that are much lower dimensional, however we require a fast way to search over thousands and potentially millions of signatures in a way that does not scale linearly with the number of sets. Locality sensitive hashing (LSH), which we describe next, is one solution to this non-exhaustive search problem.

\begin{figure}
    \centering
     \begin{subfigure}[b]{0.21\textwidth}
         \centering
         \includegraphics[width=\textwidth]{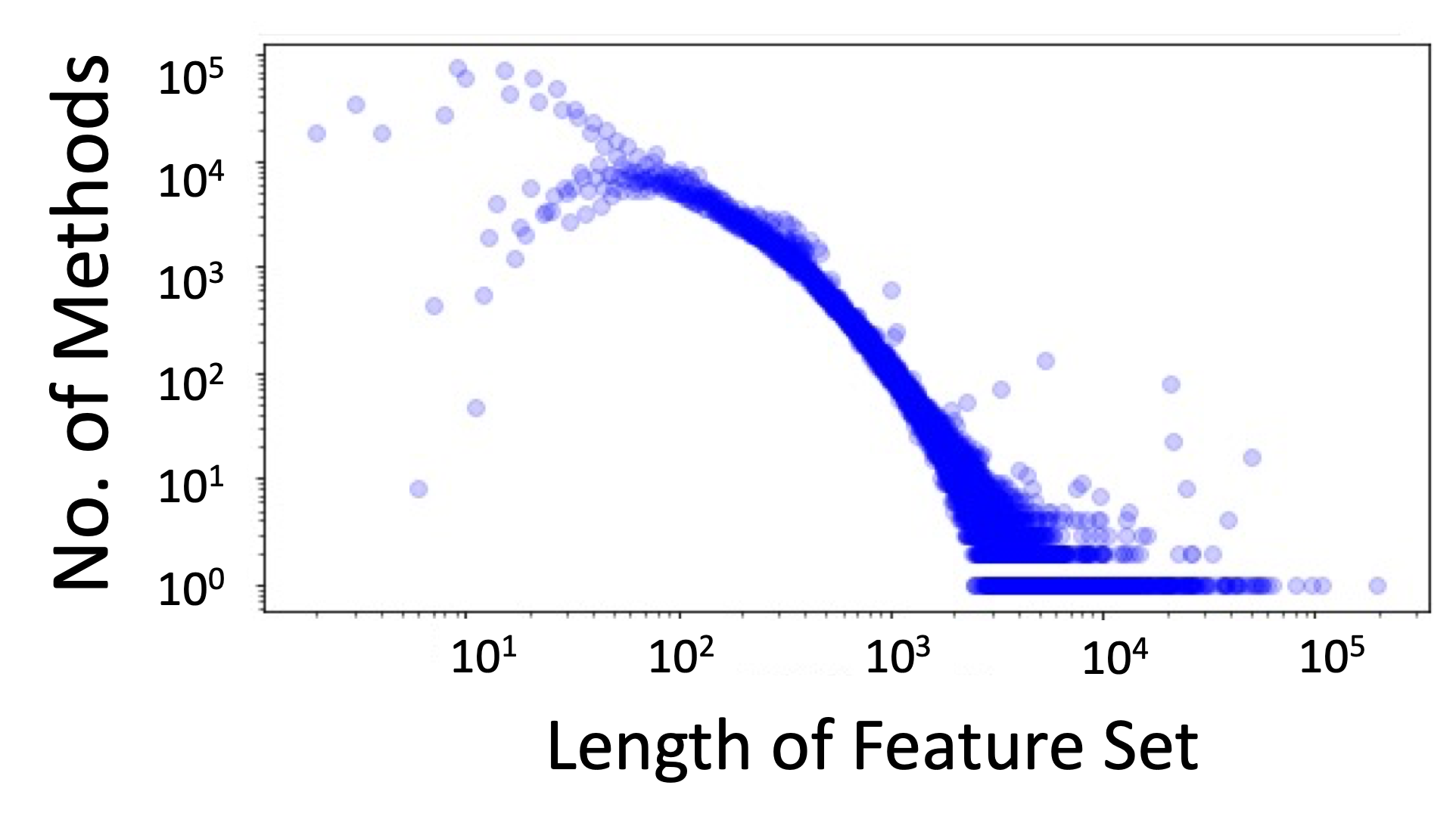}
         \caption{CodeSearchNet}
     \end{subfigure}
     \begin{subfigure}[b]{0.21\textwidth}
         \centering
         \includegraphics[width=\textwidth]{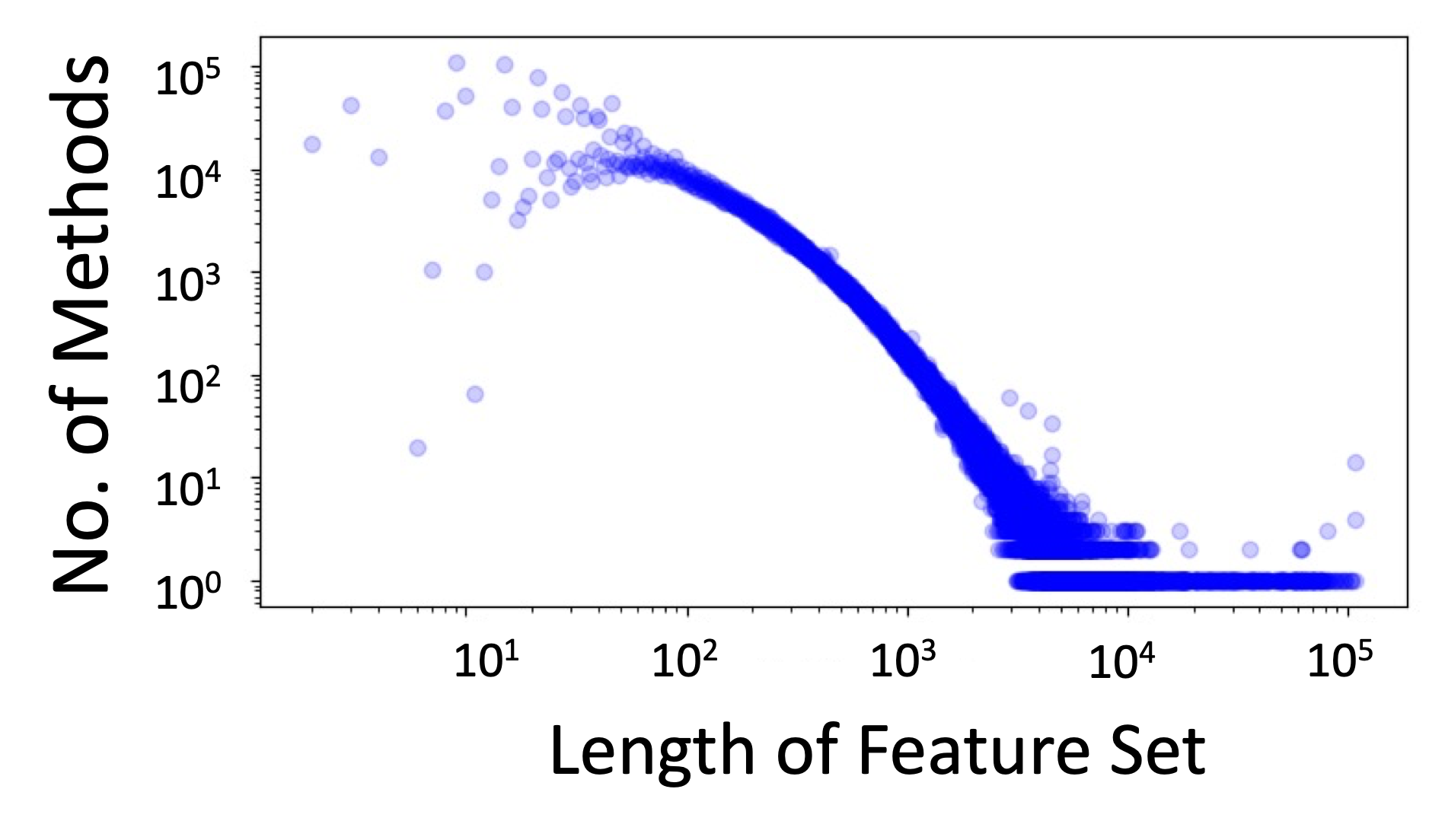}
         \caption{Neural Code Search}
     \end{subfigure}
    \caption{Distribution of Code Snippet Lengths (Log-Log Plot). We observe the characteristic linear relationship evident on a log-log plot of variables related by a power-law.}
    \label{fig:skewdist}
\end{figure}

\subsection{Locality Sensitive Hashing (LSH)}
 
In the previous section we described how a set can be converted into a signature that preserves Jaccard similarity. In this section we describe a fast technique for searching of millions of these signatures. Locality sensitive hashing (LSH)~\cite{Indyk98} is an indexing method for approximate nearest neighbour search that can facilitate sub-linear query time over that index. For large datasets and high dimension, LSH can provide compelling gains in efficiency versus brute-force search or other indexing techniques such as a kd-tree~\cite{Friedman77}. In this section we describe Minhash-LSH which is a way to construct an LSH index for minhash signatures. We note that LSH can also be applied to signatures that preserve other similarities of interest such as cosine similarity~\cite{Charikar02}, Euclidean distance~\cite{Datar04}~\etc. To index a minhash signature for set $s$, MinHash-LSH partitions the signature into $B$ bands of size $R$. $B$ is also the number of hashtables in our MinHash-LSH index.\footnote{The parameters $B$ and $R$ can be tuned to adjust the Jaccard similarity above which sets will have a high probability of colliding in a bucket.} A random hash function is applied to each band producing an integer that is the index of the colliding hash table bucket. This procedure is repeated for all $B$ bands. To retrieve similar sets from the Minhash-LSH index, the minhash signature of a query set is divided into $B$ bands and each band is hashed to produce bucket indices. These buckets are inspected and the signatures in those buckets are the candidate results for the query. To reduce false positives the Jaccard similarity of the candidates can be estimated as described in Section~\ref{sec:minhash} and those with low similarity removed from the result set. If the number of data-points in the candidate result set is much less than the dataset size, which it typically will be, then MinHash-LSH will provide a fast (sub-linear) time query efficiency.

\begin{figure*}[ht]
    \centering
    \includegraphics[width=\textwidth]{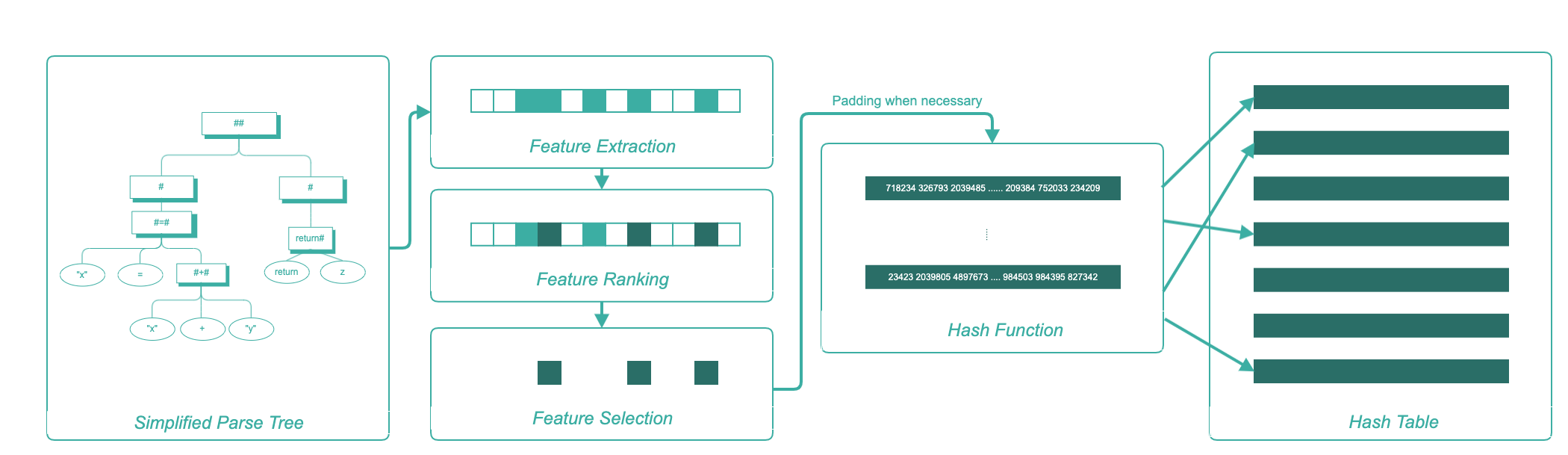}
    \caption{The core of Senatus, employing approximate similarity search via Minhash LSH to cluster similar code snippets. Left to right: simplified parse tree of a code snippet that is featurized. Feature selection (Equations~\ref{eq:featuretransform1}-\ref{eq:featuretransform2}) removes noisy features, leaving those more effective for code-to-code recommendation. Minhash signature is generated from the features and these signatures are hashed into buckets enabling sub-linear time retrieval.}
    \label{fig:overview_senatus}
\end{figure*}

\section{Methodology}\label{sec:method}

\subsection{Overview}

Similarity search is a fundamental building block for a source code retrieval system. Ever expanding codebases, such as GitHub, serve over 100 million source code repositories~\footnote{\url{https://github.blog/2018-11-08-100m-repos/\#:~:text=Today\%20we\%20reached\%20a\%20major,collaborating\%20across\%201.1\%20billion\%20contributions.}}, and this sheer scale poses challenges for efficient code search. We are in the era of ``Big Code'' a term recently coined by Allamanis~\etal~\cite{allamanis2018survey}, to describe the widespread existence of large code corpora and associated metadata (\eg commit history, comments, ratings~\etc). It is expensive to compute the exact similarity between a query and every snippet in the codebase given the linear scaling of the query-time with respect to the database size for brute force search. MinHash-LSH, described in Section~\ref{sec:definitions}, is an effective method for reducing the number of comparisons required\footnote{Inverted indexing could be used for scalable code search, however this approach has an $O(|N|^{b})$ query time, where $b$ is approximately $0.5$~\cite{Petrovic2013RealtimeED}, versus a $O(\log |N|)$ query time for LSH-based search, where $N$ is the database size.}. However, for source code we find that MinHash-LSH leads to sub-optimal retrieval effectiveness due to the heavily skewed distribution of code snippet length (Figure~\ref{fig:skewdist}). As we describe in more detail in Section~\ref{sec:deskew}, set similarity measures either favour code snippets that are shorter or  longer than the query snippet,~\eg~over 100 lines, and neither aid downstream code search and recommendation tasks. For a code-to-code recommendation to be useful, the ranked code snippets from the search engine should include the query snippet and a small number of additional lines demonstrating how that query snippet could be augmented by the developer~\cite{luan2019}. In this section we describe a new method, dubbed \emph{De-Skew LSH}, that adapts Minhash-LSH so that it can be successfully be used in the source code retrieval domain. De-Skew LSH, shown in Figure~\ref{fig:overview_senatus}, encompasses novel feature selection and pruning techniques to counteract the data skewness for MinHash-LSH. We incorporate De-Skew LSH in our code-to-code recommendation engine Senatus (Figure~\ref{fig:senatus}) and show that the engine is quantitatively better than existing approaches in our experimental evaluation in Section~\ref{sec:eval}.


\subsection{De-Skew LSH: Counteracting Data Skewness for Minhash-LSH} \label{sec:deskew}

In our empirical analysis of software repositories we find that the length of code features follows a characteristic \emph{power-law} distribution: a vast majority of code snippets are shorter with a heavy-tail of much longer length snippets. Figure \ref{fig:skewdist} demonstrates this phenomena empirically in the CodeSearchNet and Neural Code Search Evaluation datasets. Additionally, user queries tend to be shorter than the desired retrieved results, especially for code-to-code recommendation tools as developers are usually seeking ideas and suggestions for useful \emph{additional} code to augment their current function. In our experimental results (Section~\ref{sec:eval}) we present empirical evidence for the importance of appropriately handling the mismatch in snippet length in a modern code-to-code recommendation engine. 

Aroma~\cite{luan2019} defines a function $A(.)$ that featurizes a code snippet into a binary feature vector resulting from a traversal of the snippet's SPT. Aroma computes the containment score (Equation~\ref{eq:containment}) between the query snippet vector $q\in\mathbb{Z}^{D}$, and the $N$ snippet vectors $\{m_{i}\in\mathbb{Z}^{D}\}_{i=1}^{N}$ in the repository collection. The resulting scores are ranked and the snippets with a score higher than a predefined threshold are presented by Senatus to the user as relevant results for their query.

\begin{equation}
\label{eq:containment}
Containment(q, m_{i}) = \frac{|A(q) \cap A(m_{i})|}{|A(q)|}
\end{equation}

Code snippets with a longer length have a higher likelihood of overlapping with the query, thereby increasing the numerator in Equation~\ref{eq:containment}. Consequently, containment score will tend to favour snippets that are much longer in length than the query. However these longer code snippets are not necessarily relevant to the query despite the increased similarity score. In addition to potentially poor retrieval performance, users will likely have a sub-optimal experience as more time will be spent attempting to distil relevant information from the longer code snippets. Recall from Section~\ref{sec:related_work} that code-to-code recommendation tools are designed to favour snippets that usefully extend the query with a small additional amount of information.

Jaccard similarity (Equation~\eqref{eq:jaccard}) is an alternative set similarity measure that can be used to retrieve relevant code snippets. In contrast to containment score, smaller code snippets or snippets that are similar to the query length will tend to have a higher Jaccard similarity. This is a consequence of the denominator in Equation~\ref{eq:jaccard}, $|A(q) \cup A(m_{i})|$ which  penalises lengthy snippets. The bias towards shorter snippets does not bring additional value to users especially in the code-to-code recommendation task which favours retrieval results that non-trivially extend the query snippet~\ie~so that the developer can copy and paste the additional code into their function.

\useshortskip 
\begin{equation}
    \label{eq:jaccard}
    Jaccard(q, m_{i}) = \frac{|A(q) \cap A(m_{i})|}{  |A(q) \cup A(m_{i})|} 
\end{equation}

The computation of the containment score can be treated as a sparse matrix multiplication~\cite{luan2019} which means that the resulting query time can be fast, leveraging high performance libraries for numerical linear algebra. However containment score still remains a fundamentally $\mathcal{O}(|N|)$ operation, with the computation growing linearly with the dataset size $N$. In comparison, MinHash LSH, described in Section~\ref{sec:definitions} offers $\mathcal{O}(log|N|)$ retrieval time.

Asymmetric transformation~\cite{shrivastava2015asymmetric} proposes a variant of Minhash for containment score instead of the Jaccard similarity. In this approach, padding is performed to obtain vectors of the same fixed length prior to applying the hash functions. The query snippet is not padded. However this approach is sub-optimal for code-to-code recommendation: if we pad every snippet to the maximum length in the corpus, this will create a large MinHash signature as suggested by the heavy-tail in Figure \ref{fig:skewdist} and therefore will result in very high computational and memory cost.

We adapt traditional feature scoring techniques used in IR, such as the Term Frequency-Inverse Document Frequency (TF-IDF), to the source code domain. Specifically, we use a scoring function to rank the features based on relevance and only retain those within the predefined range for Minhashing. We then leverage Locality Sensitive Hashing (LSH) on these MinHash signatures, such that
the probability of collision only depends on relevant features, thus counteracting the data skewness and improving retrieval performance. The approximately similar snippets are retrieved in $\mathcal{O}(log|N|)$ time and limit the subsequent exact similarity search to only a small subset of the corpora. It is important to note that this approach is agnostic to the chosen feature selection techniques. The next section describes our feature selection methods and the application of Minhash-LSH.

\subsection{Code-to-Code Recommendation at Scale}~\label{sec:c2c}


In this section we present a methodology to score, rank and select a discriminative subset of structural sub-tree features for MinHash-LSH based on the AST representation of code.

\begin{algorithm}[t!]
	\caption{Senatus (DeSkew LSH): counteracting length bias for Minhash-LSH.}
	\label{alg:deskew}
		\begin{algorithmic}[1]
			\State $L: (length(r) | r \in records)$
			\State Predefined length parameter $maxlength \in (0, max(L)]$
			\State $T:$ Hash Table
			\State $B,R:$ Bands and rows for MinHash signature generation
			\For {each $record$ in $records$} 
				\State // Extract relevant features e.g. SPT-based features
				\State $feature \gets ExtractFeature(record)$
				\State // Apply feature transformation e.g. normalisation 
				\State $feature \gets TransformFeature(feature)$
				\State // Select the most discriminative features from the set 
				\State $feature \gets SelectFeature(feature, maxlength)$ 
				\State $l \gets$ length of $feature$
				\If {$l < maxlength$}
					\State // Not applied to the query
					\State $feature \gets Pad(feature, maxlength)$
				\EndIf 
				\State // Generate Minhash signature for feature 
				\State $s \gets MinHash(feature)$
				\State // Partition signature into $j=1\ldots B$ bands of length $R$
				\State $s_{j} \gets Chunk(s)$
				\For {$j = 1$ to $B$}
					\State // Hash signature to bucket with hash function $h$
					\State insert $feature$ in $T[j][h(s_{j})]$
				\EndFor
			\EndFor
			\State \Return $T$
		\end{algorithmic}
\end{algorithm}

In Algorithm~\ref{alg:deskew}, $ExtractFeature$ converts code snippets (\eg queries, methods or classes) to their corresponding Simplified Parse Tree (SPT). As discussed in Section~\ref{sec:spt}, the SPT representation defines the structural features in our vector space vocabulary with relationships between children and ancestors captured as independent dimensions. We propose a feature scoring and selection scheme that operates on the SPT representation and mitigates the effect of data skewness on retrieval effectiveness. Our feature scoring function is defined in Equation~\ref{eq:featuretransform1}.

\begin{equation}
    score(m,f)= 
\begin{cases}
    1, & \text{if } l_{1ower} \le S(m,f) \le l_{upper}\\
    0,              & \text{otherwise}
\end{cases}
\label{eq:featuretransform1}
\end{equation}

\begin{figure}
    \includegraphics[width=0.4\textwidth]{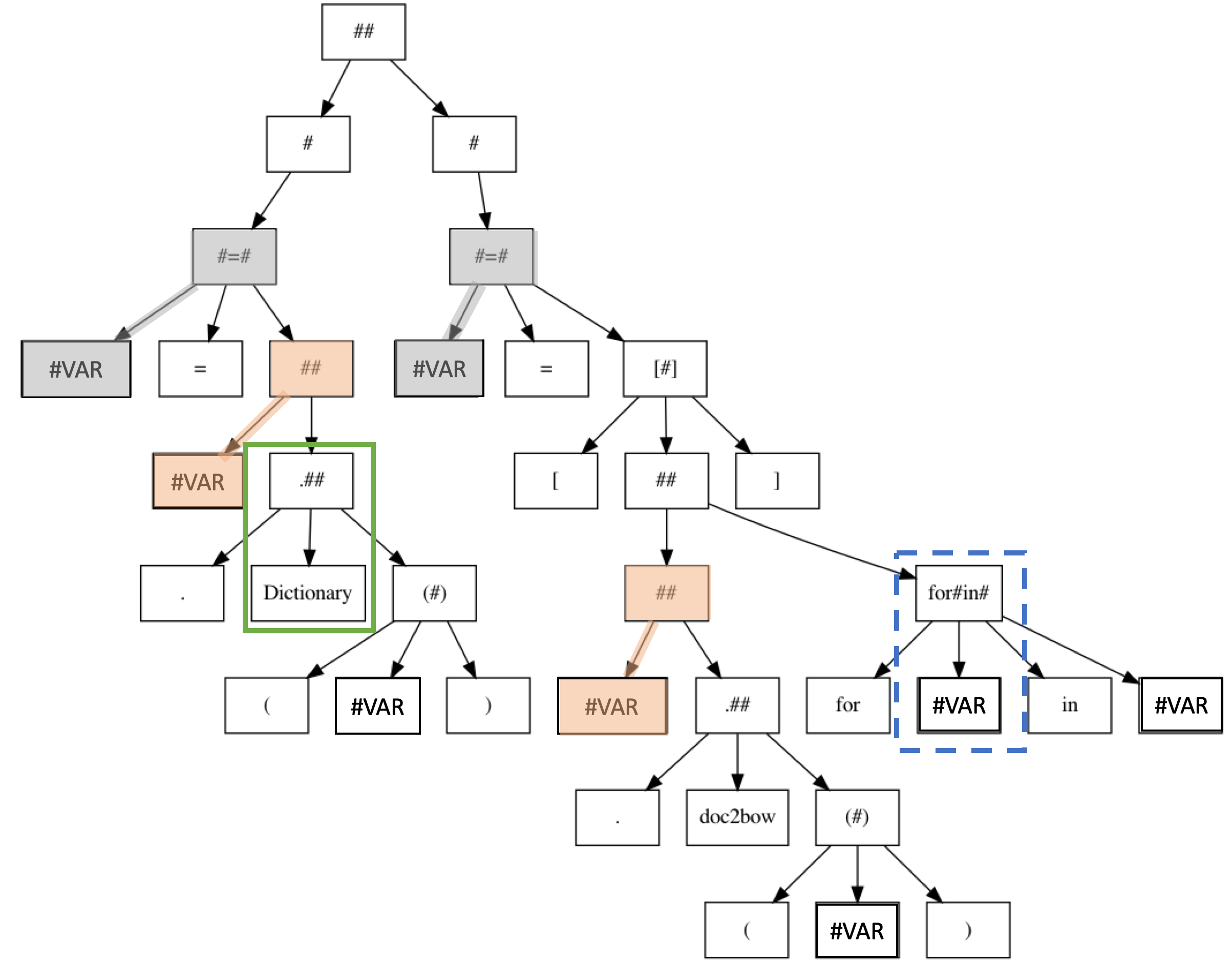}
    \caption{Comparison of structural feature scoring (Equations~\ref{eq:featuretransform2}): we propose to emphasise structural features associated with API usage (green box) with NSPF, and/or distinctive features on a method-level (blue dotted box) with ILF, instead of ordinary tree structures, such as $\#=\#1\>\#VAR$ (grey) and $\#\#1>\#VAR$ (orange).}
    \label{fig:featurescore}
\end{figure}

\noindent{In Equation~\ref{eq:featuretransform1} we define $score(m,f)$ as the score of feature $f$ with respect to the snippet $m$. The two data-driven thresholds, $l_{upper}$ and $l_{lower}$, select a subset of the terms to be used in the generation of the Minhash signatures. We retain features with a score of 1 for the subsequent minhash signature generation. We propose two different variants of the feature scoring function $S(.)$ that individually seek to emphasise different aspects of the code, for example API usage or method structures. Figure \ref{fig:featurescore} illustrates this with an example. Our proposed tree-based feature scoring functions listed in Equation~\eqref{eq:featuretransform2} rank the structural sub-tree features in the SPT representation based on their importance.}

\begin{equation}
    \label{eq:featuretransform2}
    S_{NSPF}(m,f) = \frac{|f|}{  \sum_{i=1}^{N}|f_{i}|},\hspace{0.2in} S_{ILF}(m,f) = \frac{1}{ \text{|leaf f}|} 
\end{equation}

Here $f$ refers to a particular feature e.g. \#VAR, and leaf $f$ is a leaf node feature. \emph{Normalized Sub-Path Frequency} (NSPF) is a global operator divides the frequency of tree-based features in the snippet, $m$, by the total number of occurrences of that feature in the entire codebase (of $N$ snippets), such that common features rank lower. In comparison, \emph{Inverse Leaves Frequency} (ILF) is a local operator, restricted to the query or database code snippet features only. ILF computes the inverse of the leaf node frequency within the simplified parse tree only, which removes the dependency on the background corpora and penalizes the common features on a level that is local to the given code snippet. We empirically test both variants in our experimental evaluation in Section~\ref{sec:eval}. We propose two different feature selection functions to compute the thresholds $l_{upper}$ and $l_{lower}$ in Equation~\ref{eq:featuretransform1}:

\begin{itemize}
    \item \textbf{$Top{-}F$} features where only the top $F$ highest ranking features are kept. $Top{-}F$ focuses mostly on API usage features.
    \item \textbf{$Mid{-}C$} percentile features where we rank the normalized NSPF or ILF and remove the top and bottom $(100 - C)/2$ percentile. In addition to API usage features, it focuses on method structures that typically not captured by $top{-}F$ approach. If the frequencies of the filtered features exceeds $F$, we then cap the size to $F$ by removing the bottom percentile. 
\end{itemize}

Figure~\ref{fig:selectfeature} demonstrates the selected features from an example code snippet. This is the $SelectFeature$ step in Algorithm~\ref{alg:deskew}. We shown that \textit{ILF} and \textit{NSPF} score representative features, such as API usage variables, higher, whereas standard techniques that rank purely on the count of the structural features does not, as indicated by the \textit{Full ASTs} column. Our methodology enables a similarity search to be conducted on a compact and discriminative feature set.

The final step of our proposed lightweight search process is the application of Locality Sensitive Hashing (LSH) to the MinHash signatures generated from relevant features filtered by Top-F or Mid-C percentile. With the smaller set of colliding code fragments, we apply exact similarity search by computing the dot product between the query and the candidates to further refine the results.

Our proposed DeSkew-LSH search retrieves a list of candidates ($n=100$) in sub-linear time and passes them to the second stage of the code-to-code recommendation system proposed by the Aroma authors~\cite{luan2019}. The second stage is responsible to produce the final recommendations by performing pruning and re-ranking to filter out irrelevant candidates then followed by clustering and intersecting to summarise the candidates into concise and distinct recommendations, as illustrated in Figure \ref{fig:senatus}.

\begin{figure}
    \centering
    \adjustbox{max width=0.52\textwidth}{
    \begin{lstlisting}[language=Java, basicstyle=\small]
// rank the variant graph
Set<VariantGraph.Vertex> matchedVertices = new HashSet<>();
for (List<Match> phraseMatch : phraseMatches) {
    matchedVertices.add(phraseMatch.get(0).vertex);
}
final VariantGraphRanking ranking =
VariantGraphRanking.ofOnlyCertainVertices(base, matchedVertices);
return ranking;
    \end{lstlisting}}
    \vspace{0.5cm}
    \footnotesize
    \adjustbox{max width=0.5\textwidth}{\begin{tabular}{c|c|c|c}
    \multicolumn{4}{c}{} \\
    Rank	&	Top-F + NSPF	&	Mid-95 Percentile + NSPF	&	Full ASTs	\\
    \hline 
    1	&	vertex>>VariantGraphRanking	&	\#VAR	&	\#VAR	\\
    2	&	\#.\#3>vertex	&	\#.\#1>\#VAR	&	\#.\#1>\#VAR	\\
    3	&	return\#;2>\#VAR	&	\#;1>\#VAR	&	\#;1>\#VAR	\\
    4	&	final\#\#2>VariantGraphRanking	&	\#,\#3>\#VAR	&	get>>0	\\
    5	&	0>>vertex	&	\#=\#1>>>\#,\#3	&	Vertex>>\#VAR	\\
    6	&	VariantGraphRanking	&	\#=\#1>>>return\#;2	&	\#\#:\#2>>>get	\\
    7	&	VariantGraphRanking>>\#VAR	&	\#\#:\#4>\#VAR	&	return\#;2>\#VAR	\\
    8	&	\#VAR	&	add>>>\#,\#3	&	0	\\
    9	&	\#.\#1>\#VAR	&	Vertex>>\#VAR	&	\#(\#)3>0	\\
    10	&	\#\#1>Set	&	return\#;2>\#VAR	&	\#.\#3>0	\\
     \hline
    \end{tabular}}
    \caption{Comparison in the structural features retained by the Top-F and Mid-C Percentile feature selection techniques. The example code snippet is at the top and the features retained in the table below.}
    \label{fig:selectfeature}
\end{figure}

\section{Experimental Results}\label{sec:eval}

\subsection{Datasets}\label{sec:eval2}

We evaluate retrieval effectiveness on two publicly available code search corpora: CodeSearchNet~\cite{husain2019} (CSN) and the Neural Code Search Evaluation Dataset~\cite{li2019} (NCS). The statistics of both datasets are shown in Table~\ref{table:datastat}. 

\begin{table}
    \caption{Dataset Statistics}
    \centering
    \adjustbox{max width=0.7\textwidth}{\begin{tabular}{l|c|c}
         \label{table:datastat}
          & \textbf{CodeSearchNet} & \textbf{Neural Code Search} \\ 
          \hline
         \# of repositories & 500 & 14,740 \\  
         \# of methods & 3,839,047 & 6,753,975 \\
         \# of queries & 98 & 223 \\
         Language & Java & Java \\
     \hline
    \end{tabular}}
    \label{table:datastat}
\end{table}

\textbf{Automatic evaluation:} CodeSearchNet~\cite{husain2019} consists of 99 natural language queries with their respective relevant code snippets annotated by experts with moderate agreement (Cohen's kappa of 0.47). The Neural Code Search (NCS) Evaluation Dataset~\cite{li2019} contains 287 Stack Overflow questions and code answer pairs on topics related to Android programming, where each of code answer has at least an up-vote and been manually reviewed. As we are interested in code-to-code recommendation we use the paired natural language-code snippet pairs in these datasets to generate groundtruth code-code snippet pairs. To generate our groundtruth dataset we consider as relevant snippets with the same natural language question. This permits us to measure both precision and recall of retrieval, which was not possible with the micro-benchmarking methodology employed by the Aroma authors~\cite{luan2019}. With related code snippet clusters formed in this way, we randomly choose one snippet from the cluster as the query and consider the remaining snippets as part of our database over which retrieval for that query is performed. We remove duplicate snippets by computing the SHA1 hash of the snippet's feature vector instead of a hash of the entire code snippet, where, for example, the same snippets but with different local variable naming or docstrings will be filtered out.


    \textbf{Developer Study:} In addition to the automated evaluation described above, we also conducted a developer study to evaluate retrieval effectiveness. The study is designed to mimic the real-world application of code recommendation directly in developer’s integrated development environment (IDE). In this scenario, users highlight a section of code to query for a set of recommendations. We followed the pooling procedure~\cite{jones1975report} to gather a representative selection of retrieval results to rate from the NCS dataset. More specifically, we randomly sampled a subset of 103 queries from NCS dataset, pooled and anonymised the top 10 results for those queries from three retrieval systems, one based on natural language (matching the question associated with the snippet in the NCS dataset with questions associated with the database snippets) and two based on source code structure (Aroma, Senatus). We also excluded results that were retrieved by all three systems from the evaluation to avoid duplicated results. The developers were 5 software development professionals with Masters degrees in quantitative disciplines who had on average 7.2 years of industrial programming experience and 2 years of Java programming experience. Prior to rating the developers were given detailed instructions on the task illustrated by several examples. We asked the developers to rate on a five-point Likert scale ($1{-}5$, higher is better) the \emph{usefulness} of the code recommendations in terms of extending the query snippet with relevant functionality. For each query and for each retrieval system (presented in random order), the top results (excluding duplicates) from the pool for that query and retrieval system were shown in random order to the developers for rating. In total 2,856 query-result snippet pairs were rated by the developers. The results of this study are presented in Table~\ref{table:annotation} which shows the average ratings across the 5 developers in addition to the standard deviations. We found that the developers prefer the recommendations provided by Senatus in terms of usefulness compared to the two baseline systems. 

\begin{table}
    \centering
    \caption{Mean and standard deviation averaged across developer ratings for recommendations from three search engines. The improvement to Senatus is statistically significant as compared to Aroma Lightweight Search using a paired t-test ($p < 0.05$).}
    \adjustbox{max width=0.48\textwidth}{\begin{tabular}{l|*2c}
          \hline \hline
          & \multicolumn{2}{c}{\textbf{Usefulness}} \\
          & Mean & Standard Deviation \\ 
        \hline 
         Questions Matching & 1.3674 &  1.8105 \\
         Aroma Lightweight Search & 2.2547 & 1.8084 \\  
         \textbf{Senatus DeSkew LSH} & \textbf{3.1948}* &  1.5545 \\
        \hline \hline
    \end{tabular}}
    \label{table:annotation}
\end{table}

\begin{table}
    \caption{Minhash LSH Parameter Settings: we studied the changes to retrieval performance when number of bands ($B$) and number of rows ($R$) varies with a validation set from the Neural Code Search (NCS) dataset. We then select the optimal parameters with the highest F1 score. }
    \centering
    \adjustbox{max width=\textwidth}{\begin{tabular}{ll|*5c}
        \hline \hline
        \label{table:lshparam}
B	&	R	&	Retrieval Time (s)	&	F1@100	&	Approx. Threshold	 \\
        \hline 
95	&	3	&	0.0108	&	0.3056	&	0.2191\\ 
100	&	3	&	0.0116	&	0.3052	&	0.2154\\
105	&	3	&	0.0118	&	0.3045	&	0.2119\\
95	&	2	&	0.0156	&	0.1567	&	0.1025\\
105	&	2	&	0.0184	&	0.1547	&	0.0975\\
100	&	2	&	0.0174	&	0.1564	&	0.1000\\
95	&	1	&	0.1048	&	0.1362	&	0.0105\\

     \hline \hline
    \end{tabular}}
    \label{table:lshparam}
\end{table}

\begin{table}
    \caption{Feature Selection Parameter Study: we investigated how different values of $F$ and $C$ impacts the performance and retrieval time with a validation set from Neural Code Search (NCS) dataset. The parameters that offers the highest F1 Score were selected i.e. $F=100$ and $C=95$}
    \centering
    \adjustbox{max width=\textwidth}{\begin{tabular}{ll|*5c}
        \hline \hline
        \label{table:selectparameter}
F	&	C	&	Retrieval Time (s)	&	P@100 & R@100 & F1@100 \\
        \hline 
500	&	-	&	0.0128	& 0.3278 & 0.2956	& 0.2529	\\
400	&	-	&	0.0130	& 0.3133 & 0.3310	& 0.2479	\\
300	&	-	&	0.0109	& 0.3340 & 0.4180	& 0.2839	\\
200	&	-	&	0.0110	& 0.3333 & 0.5922	& 0.3109	\\
\textbf{100}	&	-	&	0.0092	& 0.3359 & 0.7737 & \textbf{0.3349}	 \\
-	&	\textbf{95}	&	0.0116	& 0.6171 & 0.4640 & \textbf{0.4752}	& \\
-	&	96	&	0.0101	& 0.5416 & 0.4502	& 0.4347	& \\
-	&	97	&	0.0138	& 0.4630 & 0.452	& 0.3834	& \\
-	&	98	&	0.0125	& 0.4618 & 0.4929	& 0.3981	& \\
-	&	99	&	0.0129	& 0.4226 & 0.4822	& 0.3699	& \\
     \hline \hline
    \end{tabular}}
    \label{table:knparams}
\end{table}

\subsection{Hyperparameter Settings}

\begin{table*}[h!]
    \centering
    \caption{Comparison of Retrieval Performance. P, R, Time refer to the precision, recall and average query time in seconds respectively. “*" indicates that the improvement is statistically significant as compared to the baseline, Aroma, using a paired t-test ($p < 0.05$) and bold numbers indicate the best performance in that column.}
    \begin{tabular}{l|*4c|*4c}
        \hline \hline
        {} &  \multicolumn{4}{c}{\textbf{CodeSearchNet}} & \multicolumn{4}{c}{\textbf{Neural Code Search}}\\
        \hline
        {}  & P@100 & R@100 & F1@100 & Time(s) & P@100 & R@100 & F1@100 & Time(s)\\
        \hline
        Aroma Lightweight Search~\cite{luan2019}  & 0.1635  & 0.2962  & 0.1174  & 2.2771 & 0.04369 & \textbf{0.7459} & 0.0535 & 3.3599\\
        MinHash ~\cite{Broder98}   & 0.1882	& 0.3013 & 0.1375 & 0.7857* & 0.04970 & 0.05082* & 0.03939 & \textbf{0.01234*} \\
        Asymmetric MinHash~\cite{shrivastava2015asymmetric}   &  0.2042 &  \textbf{0.3039}   & 0.1496  & 2.9623 & 0.05755 & 0.7437 & 0.06870 & 2.4412*\\
        \hline
        \hline
        \textbf{Senatus} - DeSkew LSH (this paper)  &   &  &   &\\
        - Mid-C + ILF & 0.3979* & 0.2233* & 0.2233* & 0.0271* & 0.1166*	& 0.1916* &	0.0977*	& 0.0233*\\
        - Mid-C + NSPF & \textbf{0.9250}* & 0.2988 & \textbf{0.4295}* & \textbf{0.0154}* & \textbf{0.6883}* & 0.5476* & \textbf{0.5642}*	& 0.0150* \\
        - Top-F + ILF & 0.3347* & 0.2728 &  0.2096* & 0.0862* &  0.0897* & 0.7174 & 0.1037* & 0.1064* \\
        - Top-F + NSPF & 0.3678* &  0.2818  & 0.2245*  & 0.0693* & 0.1225* & 0.7316 & 0.1252* & 0.0743* \\
        \hline 
        \hline
        Corpora Size (Number of Methods) &  \multicolumn{4}{c}{2,147,445} & \multicolumn{4}{c}{2,618,306}\\
        Vocabulary Size &  \multicolumn{4}{c}{23,621,747} & \multicolumn{4}{c}{34,229,875}\\
        \hline \hline
    \end{tabular}
    \label{table:results}
\end{table*}

Minhash LSH requires the setting of the number of bands ($B$) and number of rows ($R$). We evaluated the performance and speed with varying $B$ and $R$ on 20 randomly selected queries and 200,000 code snippets from Neural Code Search (NCS) with the NSPF approach, and summarized the findings in Table \ref{table:lshparam}. We err on the side of recall and selected the range of $B$ and $R$ that gives us an approximate threshold of 0.2. The optimal parameters are $B=95, R=3$ for both NCS and CodeSearchNet (CSN).



We have also investigated the effect of $F$ and $C$ on the performance. These values determine the number of features used for MinHash, where the larger $F$ or $C$ the more features are selected and vice versa. Using the same experimental settings as the LSH parameter study, we set the values empirically via grid search: $F=100$ and $C=95$ that have the highest F1 Score. Table \ref{table:selectparameter} highlighted the most notable findings from the grid search, where $100 \leqslant F \leqslant 500$ and $95 \leqslant C \leqslant 99 $. The search range from $C$ is straight forward and the range for $F$ stems from Figure \ref{fig:skewdist}. It indicates that the majority of the code snippets are distributed between $100$ and $1,000$ for both datasets, and we err on the side of retrieval time, we studied the range between $100$ and $500$ (query time increases as the features size results increases). Therefore, we argue that the filtered number of features should also fall within the same range, as less than $100$ results in significant information loss and more than $500$ results in bias from the skewness stated in Section \ref{sec:deskew}. Both datasets share the same $F$ and $C$ values due to the close similarity in their distributions of code snippet lengths. 


\subsection{Results and Analysis}

\begin{figure}
    \centering
    \includegraphics[width=0.4\textwidth]{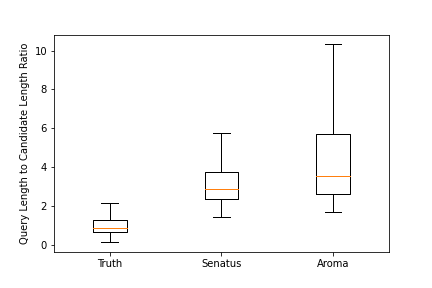}
    \adjustbox{max width=0.45\textwidth}{\begin{tabular}{l|c|c}
          \hline \hline
          & \multicolumn{2}{c}{\textbf{Similarity}} \\ 
          \hline
          & Jaccard & Containment \\
          \hline
          Senatus & 0.4581  & 1.1995 \\  
          Aroma & 0.4004  & 1.2154 \\
          Truth & 0.4805 & 1.1901 \\
     \hline \hline
    \end{tabular}}
    \caption{Comparison of the ratio of query-to-candidate snippets length and average pairwise similarity between recommended snippets. Senatus (middle) recommends more concise and diverse code snippets, whereas Aroma can recommend much longer snippets with respect to the query size, which are arguably less useful.}
    \label{table:diversity}
\end{figure}

Our quantitative evaluation examines the performance of the lightweight search stage in Aroma versus De-Skew LSH of Senatus and the end-to-end performance with the additional downstream stages of re-ranking, clustering and intersection. For Aroma lightweight search versus De-Skew LSH, Table~\ref{table:results} summarizes the quantitative results in terms of Precision@100, Recall@100 and F1@100 and average retrieval time (in seconds). Also compared are MinHash~\cite{Broder98} and Asymmetric MinHash~\cite{shrivastava2015asymmetric}. We choose @100 as \emph{only} those retrieved methods are then passed onto the downstream tasks as mentioned in Section \ref{sec:c2c}. Figure~\ref{table:diversity} summarizes the end-to-end system performance.

\begin{figure}
    \centering
    \includegraphics[width=0.5\textwidth]{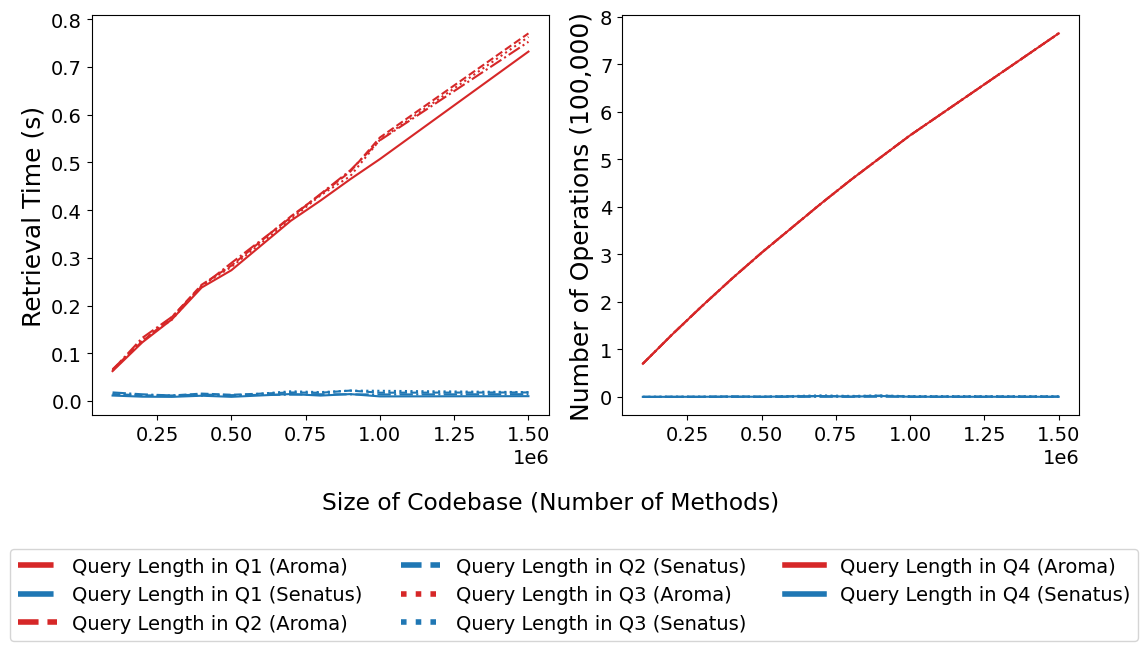}

    \caption{Comparison of Scalability in terms of the number of operations $\mathcal{O}$, retrieval time (seconds) and the impact of query length. We empirically show that Senatus (blue) has a $\mathcal{O}(log|N|)$ retrieval time regardless of query length. $Q i$ represents the $i$-th quartile, $i \in \{1,2,3,4\}$, in the distribution of feature lengths.\protect\footnotemark}
    \label{fig:scale}
\end{figure}

\footnotetext{Each quartile contains an approximately 60 queries. Q1, Q2, Q3 and Q4 range from 9 to 43, 40 to 57, 58 to 80, 81 to 352 respectively.}
    
\textbf{Improvement in retrieval time and quality:} For search over corpora with millions of methods, we find that Senatus De-Skew LSH empirically outperforms the \textit{light-weight search} stage from Aroma. In particular, on the CodeSearchNet corpus Senatus De-Skew LSH outperforms Aroma by +31.21\% F1@100 and is 147.9x faster. On the Neural Code Search dataset, Senatus De-Skew LSH outperforms Aroma by 10.55x F1@100 and is 224x faster. Comparing Senatus De-Skew LSH to the traditional MinHash approach that uses Jaccard similarity and Asymmetric MinHash we find significant increases in F1@100, showing the promise of our proposed feature scoring and feature selection methodologies (ILF, NSPF). More specifically, we found that \textit{Mid-C + NSPF} offers a better quality and speed trade-off as suggested by the performance breakdown in Table~\ref{table:results}. We further evaluated the retrieval quality of similarity against Aroma and natural language based system, where it retrieves snippets that have semantically similar questions in the NCS dataset. The semantic meaning is derived from the cosine similarity between the bag of words (BoW) representation of the natural language questions. We see a 41.7\% improvement in retrieval quality when compared to Aroma and 2.34x increase compared to question matching as indicated by the average rating in Table \ref{table:annotation}.

\textbf{Search at Scale: } Figure \ref{fig:scale} demonstrates the scalability of \textit{De-Skew LSH} stage from Senatus when compared to \textit{light-weight search} from Aroma using the NCS dataset. We measure the average number of operations and retrieval time against the number of methods in the corpus together with the impact of query length. Unlike Aroma, the retrieval time for Senatus only grows logarithmically with the corpus size and the query length has minimal impact on $\mathcal{O}$ and time. More specifically the queries with the length within the first and last quartile result in slightly shorter retrieval time as there are less colliding snippets evidenced by the power-law distributions in Figure \ref{fig:skewdist}.

\textbf{Conciseness and Diversity:} Finally, we evaluate the end-to-end performance of Aroma and Senatus with re-ranking and pruning, clustering and intersection all following the De-Skew LSH stage (see Figure~\ref{fig:senatus}). For effective code-to-code recommendation we desire that the returned snippets are both diverse and a concise extension of the query snippet. The ratio between the length of the query snippets and the ground-truth snippets is an indicator of the conciseness of the recommended snippets. Much longer (and conversely, much shorter) retrieval snippets are arguably less useful as they will likely contain functionality that is very different (and therefore not useful) to the much smaller query snippet or they will lack any additional information compared to the query snippet (if shorter than the query). In Figure~\ref{table:diversity}, the candidate lengths retrieved by Aroma are severely skewed with a heavy-tail when compared to the generated ground-truth pairs from the NCS dataset, whereas Senatus retrieves snippets closer to the groundtruth distribution. In Figure~\ref{table:diversity}, we also observe a similar behaviour with the average pairwise similarity between the recommended snippets. This measures the diversity of the recommendations, whereby a relatively lower similarity score~\ie~more distinct, leads to more diverse set of recommendations. Senatus has a lower containment score~\ie~more diverse. The bias explained in Section~\ref{sec:method} led to Senatus having a higher Jaccard similarity than Aroma as evidenced by the containment score and length ratio. The similarity between recommendations from Senatus is significantly closer to the groundtruth. These results combined provide evidence that Senatus provides more concise and diverse code recommendations.

\subsection{Threats to validity}
Threats to the validity of our findings relate to our strategy for automated evaluation and the setup of our user study. As outlined in Section~\ref{sec:eval2}, our automated evaluation harvests related code pairs from datasets of natural language-code snippet pairs by making the assumption that code snippets associated with the same natural language questions are functionally related. It can be the case that two snippets that do not share a common natural language query could still be related, thereby influencing the interpretation of our recall numbers. In our user study we ask developers to judge the usefulness of recommendations. We define usefulness specifically in terms of the code-to-code recommendation task which seeks code that can extend the code currently being written by the developer. Other definitions of usefulness are not taken into consideration, such as recommendations that are concise. Additionally we use pooling to retrieve a subset of the collection for annotation, as it is not reasonable to manually annotate every query-snippet pair. While pooling is an accepted technique in the field of Information Retrieval for building a test collection~\cite{jones1975report,Zobel98}, it can introduce bias into the evaluation particularly with larger collection sizes~\cite{Buckley06}.

\subsection{Limitations}
We highlight here that the propagation of low quality and code with security flaws can be a real risk with code-to-code recommendation tools. A practical deployment of these systems should mitigate this risk through integrated code quality checks so that only high-quality code snippets are recommended to developers.

\section{Related Work}\label{sec:related_work}

Prior art for code search and code-to-code recommendation are the closest to the work in this paper. Code search can be conducted with \emph{keyword queries}, \emph{natural language queries} or through \emph{query-by-example}~\cite{liu2020opportunities}. 

\textbf{Keyword-based Code Search:} Many code search tools tend to be text-based search engines, including OpenGrok, GitHub search, searchcode.com and Codase.\footnote{\url{https://oracle.github.io/opengrok/},\url{https://github.com/}, \url{https://searchcode.com/}, \url{http://www.codase.com/}} These source code search engines support full-text search, with options for searching for the presence of syntactic elements such as identifiers (class names, method names, attribute names) and function or variable definitions. While useful, full-text search over source code cannot leverage the syntactical structure of the code; without the structural context the retrieval may bring back many irrelevant results for the developer to sift through~\eg searching for how exception handling is performed for a file-opening command is likely to return many false positives without including surrounding structural context in the query. Commercial options for source code search include Sourcegraph, a code search and navigation engine deployed at a wide variety of large organisations. Sourcegraph offers full-text search, regex-based search and structural search via comby syntax. Extensions have sought to improve keyword-based code search~\eg by leveraging test cases~\cite{Lemos07} and using associated code documentation to improve retrieval of concise code snippets in the SNIFF search engine~\cite{Chatterjee09}.

\textbf{Natural Language-based Code Search:} Recently there has been significant interest in learning the mapping between natural language and source code~\cite{Gu18,Cambronero19}, enabling a user to search source code repositories with natural language (e.g. ``How to write data to a CSV file''). An example is Neural Code Search (NCS)~\cite{Cambronero19} that utilises neural networks to learn a shared latent space for source code and natural language. Natural language code search can be useful for certain use-cases, for example, to facilitate browsing of a repository. However, for contextually relevant code suggestions that are useful to a developer as they code~\eg fixing bugs and refactoring, natural language code search techniques fall short in convenience in comparison to query-by-example code search.

\textbf{Query-by-example code search:} In this paper we contribute new ideas to the field of query-by-example code search, and specifically in the sub-task of \emph{code-to-code recommendation}. In code-to-code recommendation, a search engine returns a list of concise, diverse and relevant code snippets given an example query code snippet as input. Code-to-code recommendation engines are based on the premise that developers typically write code that is similar to code written by other developers~\cite{luan2019}. Evidence of this usage has been provided by~\eg~Bajracharya~\etal who have analysed the log files of Koders, a web-based code search engine, and report that 90\% of the queries are related to developers searching for code to reuse in their project~\cite{Bajracharya12}. Given this, an effective method for automatically surfacing contextually relevant code snippets, portions of which could be reused, is potentially very useful in a developer's workflow. The potential benefits of an effective code-to-code recommendation engine are numerous. For example, it could ~\emph{save developer time} by presenting contextually relevant code suggestions directly in the developer's integrated development environment (IDE), or given a partially written code snippet, a developer can view suggestions that demonstrate how the snippet could be completed or extended with related functionality written by other developers. 

Prior research in the field of code search is substantial~\cite{Cordy11,Jiang07,Kamiya02,Sajani16,luan2019,li2018,bhoopchand2016learning,Cambronero19,Balachandran15,Shuhan20,deRezende20,Wei20,Krugler2013,Kim18}, with the recent review article summarising the research~\cite{liu2020opportunities}. Early work in this field includes~\emph{clone detection} which are systems that aim to find highly similar code snippets known as Type 1-3 code clones (\ie syntactically similar code). Relevant prior work in this area includes CCFinder and SourcererCC~\cite{Kamiya02,Sajani16}. Recent work in clone detection has Type 4 clones (semantically similar code)~\cite{White16,Saini18} and gapped clones~\cite{Ueda02}. As argued in~\cite{luan2019}, clone detectors are not suitable for the task of code-to-code recommendation, where the desire is for the retrieved code snippets to contain both the query and additional useful code that could be re-purposed by the developer for their task. Other related work includes code-to-code search tools such as FaCoY~\cite{Kim18} and Krugle~\cite{Krugler2013},\footnote{\url{https://opensearch.krugle.org/}} however, again these type of tools are not useful for code-to-code recommendation as they may produce redundant results in the ranked list with extraneous additional code in the result snippets that are not a concise extension of the query.

{\textbf{Facebook Aroma:}} Aroma~\cite{luan2019} provides diverse and concise retrieval results that are a useful extension of the query. Aroma is a structured code-to-code recommendation engine and at its core is a featurisation and similarity computation algorithm based on a \emph{Simplified Abstract Parse Tree (SPT)} representation that facilitates recommendation by tolerating a degree of non-similarity in very similar snippets. The Aroma Abstract Syntax Tree (AST) is simplified into a SPT by replacing local variable names with a common tag (\eg \#VAR) unless they are global variables or method names, as they can be important in differentiating code snippets. In order to capture structural relationships, Aroma further creates features using three types of relationships between variable-based tokens on the AST: a) parent, b) sibling and c) variable re-use. The individual tokens as well as the structural relationship features parsed from the code corpus constitute the \emph{vocabulary} for the task, and are given individual indices which are then used to convert the AST into a vector. Aroma accepts a partial query snippet and searches for method bodies which contain that snippet. In order to achieve this quickly, Aroma first performs a fast light-weight search on the whole source code corpus, followed by slower pruning, re-ranking and intersection (clustering) processes on the selected (\eg~1000) results that produce a diverse and concise result set.

\textbf{Minwise Hashing:} advancements in minwise hashing (MinHash) are relevant to Senatus data indexing and query processing. MinHash was introduced in the seminal papers by Broder~\etal~\cite{Broder98,Broder97} and converts sets to signatures that preserve the Jaccard similarity of the sets. Locality sensitive hashing (LSH)~\cite{Andoni08,Indyk98} can be applied to these signatures so that similar signatures collide in the same hashtable buckets with high probability. MinHash has subsequently been extended in several directions to improve its performance and efficiency: for example the method of~\cite{shrivastava2015asymmetric,Zhu16} counteracts bias towards smaller sets, Ioffe~\etal~\cite{Ioffe10} extend MinHash to weighted Jaccard similarity on multisets and b-bit MinHash~\cite{Li10} targets storage space savings by retaining only the lowest b-bits of each hash value. 

\section{Conclusions}\label{sec:conclusions}

In this paper we presented \emph{Senatus}, a new code-to-code recommendation engine that empirically demonstrates high scalability and best-in-class accuracy. The search engine was also preferred to competitive baselines by a group of developers. Underlying Senatus is our novel proposal for a new variant of MinHash-LSH which we call \emph{De-Skew LSH}. Minhash-LSH is a well-known technique for achieving a sub-linear time retrieval over data that is represented as sets. Minhash-LSH is negatively influenced by the highly skewed distribution of code snippet lengths found in typical repositories. We propose a new approximate nearest neighbour search algorithm De-Skew LSH that counteracts this skewness and improves retrieval effectiveness without sacrificing retrieval time.
De-Skew LSH applies novel feature scoring (Inverse Leaves Frequency, Normalized Sub-Path Frequency) and selection (Top-F, Mid-C percentile) methods to the abstract syntax tree representation of code to equalise the length of snippets and focus on the most discriminative feature set for retrieval.

Future work will explore learning-to-rank for improving Senatus recommendations potentially using weakly supervised labels to reduce the annotation burden~\cite{Zamani18}. Furthermore, we plan to extend Senatus for Jupyter Notebook code recommendation which we believe will be of significant value for data science workflows. 


\bibliographystyle{ACM-Reference-Format}
\bibliography{sample-base}


\begin{thebibliography}{59}


\ifx \showCODEN    \undefined \def \showCODEN     #1{\unskip}     \fi
\ifx \showDOI      \undefined \def \showDOI       #1{#1}\fi
\ifx \showISBNx    \undefined \def \showISBNx     #1{\unskip}     \fi
\ifx \showISBNxiii \undefined \def \showISBNxiii  #1{\unskip}     \fi
\ifx \showISSN     \undefined \def \showISSN      #1{\unskip}     \fi
\ifx \showLCCN     \undefined \def \showLCCN      #1{\unskip}     \fi
\ifx \shownote     \undefined \def \shownote      #1{#1}          \fi
\ifx \showarticletitle \undefined \def \showarticletitle #1{#1}   \fi
\ifx \showURL      \undefined \def \showURL       {\relax}        \fi
\providecommand\bibfield[2]{#2}
\providecommand\bibinfo[2]{#2}
\providecommand\natexlab[1]{#1}
\providecommand\showeprint[2][]{arXiv:#2}

\bibitem[\protect\citeauthoryear{Allamanis, Barr, Devanbu, and
  Sutton}{Allamanis et~al\mbox{.}}{2018}]%
        {allamanis2018survey}
\bibfield{author}{\bibinfo{person}{Miltiadis Allamanis},
  \bibinfo{person}{Earl~T Barr}, \bibinfo{person}{Premkumar Devanbu}, {and}
  \bibinfo{person}{Charles Sutton}.} \bibinfo{year}{2018}\natexlab{}.
\newblock \showarticletitle{A survey of machine learning for big code and
  naturalness}.
\newblock \bibinfo{journal}{\emph{ACM Computing Surveys (CSUR)}}
  \bibinfo{volume}{51}, \bibinfo{number}{4} (\bibinfo{year}{2018}),
  \bibinfo{pages}{81}.
\newblock


\bibitem[\protect\citeauthoryear{Alon, Zilberstein, Levy, and Yahav}{Alon
  et~al\mbox{.}}{2019}]%
        {Alon19}
\bibfield{author}{\bibinfo{person}{Uri Alon}, \bibinfo{person}{Meital
  Zilberstein}, \bibinfo{person}{Omer Levy}, {and} \bibinfo{person}{Eran
  Yahav}.} \bibinfo{year}{2019}\natexlab{}.
\newblock \showarticletitle{Code2vec: Learning Distributed Representations of
  Code}.
\newblock \bibinfo{journal}{\emph{Proc. ACM Program. Lang.}}
  \bibinfo{volume}{3}, \bibinfo{number}{POPL}, Article \bibinfo{articleno}{40}
  (\bibinfo{date}{jan} \bibinfo{year}{2019}), \bibinfo{numpages}{29}~pages.
\newblock
\urldef\tempurl%
\url{https://doi.org/10.1145/3290353}
\showDOI{\tempurl}


\bibitem[\protect\citeauthoryear{Andoni and Indyk}{Andoni and Indyk}{2008}]%
        {Andoni08}
\bibfield{author}{\bibinfo{person}{Alexandr Andoni} {and}
  \bibinfo{person}{Piotr Indyk}.} \bibinfo{year}{2008}\natexlab{}.
\newblock \showarticletitle{Near-Optimal Hashing Algorithms for Approximate
  Nearest Neighbor in High Dimensions}.
\newblock \bibinfo{journal}{\emph{Commun. ACM}} \bibinfo{volume}{51},
  \bibinfo{number}{1} (\bibinfo{date}{Jan.} \bibinfo{year}{2008}),
  \bibinfo{pages}{117–122}.
\newblock
\showISSN{0001-0782}
\urldef\tempurl%
\url{https://doi.org/10.1145/1327452.1327494}
\showDOI{\tempurl}


\bibitem[\protect\citeauthoryear{Bajracharya and Lopes}{Bajracharya and
  Lopes}{2012}]%
        {Bajracharya12}
\bibfield{author}{\bibinfo{person}{Sushil~Krishna Bajracharya} {and}
  \bibinfo{person}{Cristina~Videira Lopes}.} \bibinfo{year}{2012}\natexlab{}.
\newblock \showarticletitle{Analyzing and Mining a Code Search Engine Usage
  Log}.
\newblock \bibinfo{journal}{\emph{Empirical Softw. Engg.}}
  \bibinfo{volume}{17}, \bibinfo{number}{4–5} (\bibinfo{date}{Aug.}
  \bibinfo{year}{2012}), \bibinfo{pages}{424–466}.
\newblock
\showISSN{1382-3256}
\urldef\tempurl%
\url{https://doi.org/10.1007/s10664-010-9144-6}
\showDOI{\tempurl}


\bibitem[\protect\citeauthoryear{Balachandran}{Balachandran}{2015}]%
        {Balachandran15}
\bibfield{author}{\bibinfo{person}{Vipin Balachandran}.}
  \bibinfo{year}{2015}\natexlab{}.
\newblock \showarticletitle{Query by example in large-scale code repositories}.
  In \bibinfo{booktitle}{\emph{2015 {IEEE} International Conference on Software
  Maintenance and Evolution, {ICSME} 2015, Bremen, Germany, September 29 -
  October 1, 2015}}, \bibfield{editor}{\bibinfo{person}{Rainer Koschke},
  \bibinfo{person}{Jens Krinke}, {and} \bibinfo{person}{Martin~P. Robillard}}
  (Eds.). \bibinfo{publisher}{{IEEE} Computer Society},
  \bibinfo{pages}{467--476}.
\newblock
\urldef\tempurl%
\url{https://doi.org/10.1109/ICSM.2015.7332498}
\showDOI{\tempurl}


\bibitem[\protect\citeauthoryear{Bhoopchand, Rocktäschel, Barr, and
  Riedel}{Bhoopchand et~al\mbox{.}}{2016}]%
        {bhoopchand2016learning}
\bibfield{author}{\bibinfo{person}{Avishkar Bhoopchand}, \bibinfo{person}{Tim
  Rocktäschel}, \bibinfo{person}{Earl Barr}, {and} \bibinfo{person}{Sebastian
  Riedel}.} \bibinfo{year}{2016}\natexlab{}.
\newblock \bibinfo{title}{Learning Python Code Suggestion with a Sparse Pointer
  Network}.
\newblock
\newblock
\showeprint[arxiv]{1611.08307}~[cs.NE]


\bibitem[\protect\citeauthoryear{Broder}{Broder}{1997}]%
        {Broder97}
\bibfield{author}{\bibinfo{person}{A.Z. Broder}.}
  \bibinfo{year}{1997}\natexlab{}.
\newblock \showarticletitle{On the resemblance and containment of documents}.
  In \bibinfo{booktitle}{\emph{Proceedings. Compression and Complexity of
  SEQUENCES 1997 (Cat. No.97TB100171)}}. \bibinfo{pages}{21--29}.
\newblock
\urldef\tempurl%
\url{https://doi.org/10.1109/SEQUEN.1997.666900}
\showDOI{\tempurl}


\bibitem[\protect\citeauthoryear{Broder, Charikar, Frieze, and
  Mitzenmacher}{Broder et~al\mbox{.}}{1998}]%
        {Broder98}
\bibfield{author}{\bibinfo{person}{Andrei~Z. Broder}, \bibinfo{person}{Moses
  Charikar}, \bibinfo{person}{Alan~M. Frieze}, {and} \bibinfo{person}{Michael
  Mitzenmacher}.} \bibinfo{year}{1998}\natexlab{}.
\newblock \showarticletitle{Min-Wise Independent Permutations (Extended
  Abstract)}. In \bibinfo{booktitle}{\emph{Proceedings of the Thirtieth Annual
  ACM Symposium on Theory of Computing}} (Dallas, Texas, USA)
  \emph{(\bibinfo{series}{STOC '98})}. \bibinfo{publisher}{Association for
  Computing Machinery}, \bibinfo{address}{New York, NY, USA},
  \bibinfo{pages}{327–336}.
\newblock
\showISBNx{0897919629}
\urldef\tempurl%
\url{https://doi.org/10.1145/276698.276781}
\showDOI{\tempurl}


\bibitem[\protect\citeauthoryear{Buckley, Dimmick, Soboroff, and
  Voorhees}{Buckley et~al\mbox{.}}{2006}]%
        {Buckley06}
\bibfield{author}{\bibinfo{person}{Chris Buckley}, \bibinfo{person}{Darrin
  Dimmick}, \bibinfo{person}{Ian Soboroff}, {and} \bibinfo{person}{Ellen
  Voorhees}.} \bibinfo{year}{2006}\natexlab{}.
\newblock \showarticletitle{Bias and the Limits of Pooling}. In
  \bibinfo{booktitle}{\emph{Proceedings of the 29th Annual International ACM
  SIGIR Conference on Research and Development in Information Retrieval}}
  (Seattle, Washington, USA) \emph{(\bibinfo{series}{SIGIR '06})}.
  \bibinfo{publisher}{Association for Computing Machinery},
  \bibinfo{address}{New York, NY, USA}, \bibinfo{pages}{619–620}.
\newblock
\showISBNx{1595933697}


\bibitem[\protect\citeauthoryear{Cambronero, Li, Kim, Sen, and
  Chandra}{Cambronero et~al\mbox{.}}{2019}]%
        {Cambronero19}
\bibfield{author}{\bibinfo{person}{Jose Cambronero}, \bibinfo{person}{Hongyu
  Li}, \bibinfo{person}{Seohyun Kim}, \bibinfo{person}{Koushik Sen}, {and}
  \bibinfo{person}{Satish Chandra}.} \bibinfo{year}{2019}\natexlab{}.
\newblock \showarticletitle{When Deep Learning Met Code Search}. In
  \bibinfo{booktitle}{\emph{Proceedings of the 2019 27th ACM Joint Meeting on
  European Software Engineering Conference and Symposium on the Foundations of
  Software Engineering}} (Tallinn, Estonia) \emph{(\bibinfo{series}{ESEC/FSE
  2019})}. \bibinfo{publisher}{Association for Computing Machinery},
  \bibinfo{address}{New York, NY, USA}, \bibinfo{pages}{964–974}.
\newblock
\showISBNx{9781450355728}
\urldef\tempurl%
\url{https://doi.org/10.1145/3338906.3340458}
\showDOI{\tempurl}


\bibitem[\protect\citeauthoryear{Charikar}{Charikar}{2002}]%
        {Charikar02}
\bibfield{author}{\bibinfo{person}{Moses~S. Charikar}.}
  \bibinfo{year}{2002}\natexlab{}.
\newblock \showarticletitle{Similarity Estimation Techniques from Rounding
  Algorithms}. In \bibinfo{booktitle}{\emph{Proceedings of the Thiry-Fourth
  Annual ACM Symposium on Theory of Computing}} (Montreal, Quebec, Canada)
  \emph{(\bibinfo{series}{STOC '02})}. \bibinfo{publisher}{Association for
  Computing Machinery}, \bibinfo{address}{New York, NY, USA},
  \bibinfo{pages}{380–388}.
\newblock
\showISBNx{1581134959}
\urldef\tempurl%
\url{https://doi.org/10.1145/509907.509965}
\showDOI{\tempurl}


\bibitem[\protect\citeauthoryear{Chatterjee, Juvekar, and Sen}{Chatterjee
  et~al\mbox{.}}{2009}]%
        {Chatterjee09}
\bibfield{author}{\bibinfo{person}{Shaunak Chatterjee}, \bibinfo{person}{Sudeep
  Juvekar}, {and} \bibinfo{person}{Koushik Sen}.}
  \bibinfo{year}{2009}\natexlab{}.
\newblock \showarticletitle{SNIFF: A Search Engine for Java Using Free-Form
  Queries}. In \bibinfo{booktitle}{\emph{Proceedings of the 12th International
  Conference on Fundamental Approaches to Software Engineering: Held as Part of
  the Joint European Conferences on Theory and Practice of Software, ETAPS
  2009}} (York, UK) \emph{(\bibinfo{series}{FASE '09})}.
  \bibinfo{publisher}{Springer-Verlag}, \bibinfo{address}{Berlin, Heidelberg},
  \bibinfo{pages}{385–400}.
\newblock
\showISBNx{9783642005923}
\urldef\tempurl%
\url{https://doi.org/10.1007/978-3-642-00593-0_26}
\showDOI{\tempurl}


\bibitem[\protect\citeauthoryear{Cordy and Roy}{Cordy and Roy}{2011}]%
        {Cordy11}
\bibfield{author}{\bibinfo{person}{James~R. Cordy} {and}
  \bibinfo{person}{Chanchal~K. Roy}.} \bibinfo{year}{2011}\natexlab{}.
\newblock \showarticletitle{The NiCad Clone Detector}. In
  \bibinfo{booktitle}{\emph{2011 IEEE 19th International Conference on Program
  Comprehension}}. \bibinfo{pages}{219--220}.
\newblock
\urldef\tempurl%
\url{https://doi.org/10.1109/ICPC.2011.26}
\showDOI{\tempurl}


\bibitem[\protect\citeauthoryear{Cormode and Muthukrishnan}{Cormode and
  Muthukrishnan}{2005}]%
        {Cormode05}
\bibfield{author}{\bibinfo{person}{Graham Cormode} {and} \bibinfo{person}{S.
  Muthukrishnan}.} \bibinfo{year}{2005}\natexlab{}.
\newblock \showarticletitle{Space Efficient Mining of Multigraph Streams}. In
  \bibinfo{booktitle}{\emph{Proceedings of the Twenty-Fourth ACM
  SIGMOD-SIGACT-SIGART Symposium on Principles of Database Systems}}
  (Baltimore, Maryland) \emph{(\bibinfo{series}{PODS '05})}.
  \bibinfo{publisher}{Association for Computing Machinery},
  \bibinfo{address}{New York, NY, USA}, \bibinfo{pages}{271–282}.
\newblock


\bibitem[\protect\citeauthoryear{Das, Datar, Garg, and Rajaram}{Das
  et~al\mbox{.}}{2007}]%
        {Das07}
\bibfield{author}{\bibinfo{person}{Abhinandan~S. Das}, \bibinfo{person}{Mayur
  Datar}, \bibinfo{person}{Ashutosh Garg}, {and} \bibinfo{person}{Shyam
  Rajaram}.} \bibinfo{year}{2007}\natexlab{}.
\newblock \showarticletitle{Google News Personalization: Scalable Online
  Collaborative Filtering}. In \bibinfo{booktitle}{\emph{Proceedings of the
  16th International Conference on World Wide Web}} (Banff, Alberta, Canada)
  \emph{(\bibinfo{series}{WWW '07})}. \bibinfo{publisher}{Association for
  Computing Machinery}, \bibinfo{address}{New York, NY, USA},
  \bibinfo{pages}{271–280}.
\newblock
\showISBNx{9781595936547}
\urldef\tempurl%
\url{https://doi.org/10.1145/1242572.1242610}
\showDOI{\tempurl}


\bibitem[\protect\citeauthoryear{Datar, Immorlica, Indyk, and Mirrokni}{Datar
  et~al\mbox{.}}{2004}]%
        {Datar04}
\bibfield{author}{\bibinfo{person}{Mayur Datar}, \bibinfo{person}{Nicole
  Immorlica}, \bibinfo{person}{Piotr Indyk}, {and} \bibinfo{person}{Vahab~S.
  Mirrokni}.} \bibinfo{year}{2004}\natexlab{}.
\newblock \showarticletitle{Locality-Sensitive Hashing Scheme Based on p-Stable
  Distributions}. In \bibinfo{booktitle}{\emph{Proceedings of the Twentieth
  Annual Symposium on Computational Geometry}} (Brooklyn, New York, USA)
  \emph{(\bibinfo{series}{SCG '04})}. \bibinfo{publisher}{Association for
  Computing Machinery}, \bibinfo{address}{New York, NY, USA},
  \bibinfo{pages}{253–262}.
\newblock
\showISBNx{1581138857}
\urldef\tempurl%
\url{https://doi.org/10.1145/997817.997857}
\showDOI{\tempurl}


\bibitem[\protect\citeauthoryear{de~Rezende~Martins and
  Gerosa}{de~Rezende~Martins and Gerosa}{2020}]%
        {deRezende20}
\bibfield{author}{\bibinfo{person}{Marcelo de Rezende~Martins} {and}
  \bibinfo{person}{Marco~Aur\'{e}lio Gerosa}.} \bibinfo{year}{2020}\natexlab{}.
\newblock \showarticletitle{CoNCRA: A Convolutional Neural Networks Code
  Retrieval Approach}. In \bibinfo{booktitle}{\emph{Proceedings of the 34th
  Brazilian Symposium on Software Engineering}} (Natal, Brazil)
  \emph{(\bibinfo{series}{SBES '20})}. \bibinfo{publisher}{Association for
  Computing Machinery}, \bibinfo{address}{New York, NY, USA},
  \bibinfo{pages}{526–531}.
\newblock
\showISBNx{9781450387538}
\urldef\tempurl%
\url{https://doi.org/10.1145/3422392.3422462}
\showDOI{\tempurl}


\bibitem[\protect\citeauthoryear{Friedman, Bentley, and Finkel}{Friedman
  et~al\mbox{.}}{1977}]%
        {Friedman77}
\bibfield{author}{\bibinfo{person}{Jerome~H. Friedman},
  \bibinfo{person}{Jon~Louis Bentley}, {and} \bibinfo{person}{Raphael~Ari
  Finkel}.} \bibinfo{year}{1977}\natexlab{}.
\newblock \showarticletitle{An Algorithm for Finding Best Matches in
  Logarithmic Expected Time}.
\newblock \bibinfo{journal}{\emph{ACM Trans. Math. Softw.}}
  \bibinfo{volume}{3}, \bibinfo{number}{3} (\bibinfo{date}{sep}
  \bibinfo{year}{1977}), \bibinfo{pages}{209–226}.
\newblock
\showISSN{0098-3500}
\urldef\tempurl%
\url{https://doi.org/10.1145/355744.355745}
\showDOI{\tempurl}


\bibitem[\protect\citeauthoryear{{Gu}, {Zhang}, and {Kim}}{{Gu}
  et~al\mbox{.}}{2018}]%
        {Gu18}
\bibfield{author}{\bibinfo{person}{X. {Gu}}, \bibinfo{person}{H. {Zhang}},
  {and} \bibinfo{person}{S. {Kim}}.} \bibinfo{year}{2018}\natexlab{}.
\newblock \showarticletitle{Deep Code Search}. In
  \bibinfo{booktitle}{\emph{2018 IEEE/ACM 40th International Conference on
  Software Engineering (ICSE)}}. \bibinfo{pages}{933--944}.
\newblock
\urldef\tempurl%
\url{https://doi.org/10.1145/3180155.3180167}
\showDOI{\tempurl}


\bibitem[\protect\citeauthoryear{Husain, Wu, Gazit, Allamanis, and
  Brockschmidt}{Husain et~al\mbox{.}}{2019}]%
        {husain2019}
\bibfield{author}{\bibinfo{person}{Hamel Husain}, \bibinfo{person}{Ho{-}Hsiang
  Wu}, \bibinfo{person}{Tiferet Gazit}, \bibinfo{person}{Miltiadis Allamanis},
  {and} \bibinfo{person}{Marc Brockschmidt}.} \bibinfo{year}{2019}\natexlab{}.
\newblock \showarticletitle{CodeSearchNet Challenge: Evaluating the State of
  Semantic Code Search}.
\newblock \bibinfo{journal}{\emph{CoRR}}  \bibinfo{volume}{abs/1909.09436}
  (\bibinfo{year}{2019}).
\newblock


\bibitem[\protect\citeauthoryear{Husain, Wu, Gazit, Allamanis, and
  Brockschmidt}{Husain et~al\mbox{.}}{2020}]%
        {husain2020codesearchnet}
\bibfield{author}{\bibinfo{person}{Hamel Husain}, \bibinfo{person}{Ho-Hsiang
  Wu}, \bibinfo{person}{Tiferet Gazit}, \bibinfo{person}{Miltiadis Allamanis},
  {and} \bibinfo{person}{Marc Brockschmidt}.} \bibinfo{year}{2020}\natexlab{}.
\newblock \bibinfo{title}{CodeSearchNet Challenge: Evaluating the State of
  Semantic Code Search}.  (\bibinfo{date}{June} \bibinfo{year}{2020}).
\newblock


\bibitem[\protect\citeauthoryear{Indyk and Motwani}{Indyk and Motwani}{1998}]%
        {Indyk98}
\bibfield{author}{\bibinfo{person}{Piotr Indyk} {and} \bibinfo{person}{Rajeev
  Motwani}.} \bibinfo{year}{1998}\natexlab{}.
\newblock \showarticletitle{Approximate Nearest Neighbors: Towards Removing the
  Curse of Dimensionality}. In \bibinfo{booktitle}{\emph{Proceedings of the
  Thirtieth Annual ACM Symposium on Theory of Computing}} (Dallas, Texas, USA)
  \emph{(\bibinfo{series}{STOC '98})}. \bibinfo{publisher}{Association for
  Computing Machinery}, \bibinfo{address}{New York, NY, USA}.
\newblock


\bibitem[\protect\citeauthoryear{Ioffe}{Ioffe}{2010}]%
        {Ioffe10}
\bibfield{author}{\bibinfo{person}{Sergey Ioffe}.}
  \bibinfo{year}{2010}\natexlab{}.
\newblock \showarticletitle{Improved Consistent Sampling, Weighted Minhash and
  L1 Sketching} \emph{(\bibinfo{series}{ICDM '10})}. \bibinfo{publisher}{IEEE
  Computer Society}, \bibinfo{address}{USA}, \bibinfo{pages}{246–255}.
\newblock
\showISBNx{9780769542560}


\bibitem[\protect\citeauthoryear{Jiang, Misherghi, Su, and Glondu}{Jiang
  et~al\mbox{.}}{2007}]%
        {Jiang07}
\bibfield{author}{\bibinfo{person}{Lingxiao Jiang}, \bibinfo{person}{Ghassan
  Misherghi}, \bibinfo{person}{Zhendong Su}, {and} \bibinfo{person}{Stephane
  Glondu}.} \bibinfo{year}{2007}\natexlab{}.
\newblock \showarticletitle{DECKARD: Scalable and Accurate Tree-Based Detection
  of Code Clones}. In \bibinfo{booktitle}{\emph{29th International Conference
  on Software Engineering (ICSE'07)}}. \bibinfo{pages}{96--105}.
\newblock
\urldef\tempurl%
\url{https://doi.org/10.1109/ICSE.2007.30}
\showDOI{\tempurl}


\bibitem[\protect\citeauthoryear{Jones, Van~Rijsbergen, Research, and
  Department}{Jones et~al\mbox{.}}{1975}]%
        {jones1975report}
\bibfield{author}{\bibinfo{person}{K.S. Jones}, \bibinfo{person}{C.J.
  Van~Rijsbergen}, \bibinfo{person}{British~Library. Research}, {and}
  \bibinfo{person}{Development Department}.} \bibinfo{year}{1975}\natexlab{}.
\newblock \bibinfo{booktitle}{\emph{Report on the Need for and Provision of an
  'ideal' Information Retrieval Test Collection}}.
\newblock \bibinfo{publisher}{University Computer Laboratory}.
\newblock
\urldef\tempurl%
\url{https://books.google.co.uk/books?id=cuGnSgAACAAJ}
\showURL{%
\tempurl}


\bibitem[\protect\citeauthoryear{{Kamiya}, {Kusumoto}, and {Inoue}}{{Kamiya}
  et~al\mbox{.}}{2002}]%
        {Kamiya02}
\bibfield{author}{\bibinfo{person}{T. {Kamiya}}, \bibinfo{person}{S.
  {Kusumoto}}, {and} \bibinfo{person}{K. {Inoue}}.}
  \bibinfo{year}{2002}\natexlab{}.
\newblock \showarticletitle{CCFinder: a multilinguistic token-based code clone
  detection system for large scale source code}.
\newblock \bibinfo{journal}{\emph{IEEE Transactions on Software Engineering}}
  \bibinfo{volume}{28}, \bibinfo{number}{7} (\bibinfo{year}{2002}),
  \bibinfo{pages}{654--670}.
\newblock
\urldef\tempurl%
\url{https://doi.org/10.1109/TSE.2002.1019480}
\showDOI{\tempurl}


\bibitem[\protect\citeauthoryear{Kim, Kim, Bissyand\'{e}, Choi, Li, Klein, and
  Traon}{Kim et~al\mbox{.}}{2018}]%
        {Kim18}
\bibfield{author}{\bibinfo{person}{Kisub Kim}, \bibinfo{person}{Dongsun Kim},
  \bibinfo{person}{Tegawend\'{e}~F. Bissyand\'{e}}, \bibinfo{person}{Eunjong
  Choi}, \bibinfo{person}{Li Li}, \bibinfo{person}{Jacques Klein}, {and}
  \bibinfo{person}{Yves~Le Traon}.} \bibinfo{year}{2018}\natexlab{}.
\newblock \showarticletitle{FaCoY: A Code-to-Code Search Engine}. In
  \bibinfo{booktitle}{\emph{Proceedings of the 40th International Conference on
  Software Engineering}} (Gothenburg, Sweden) \emph{(\bibinfo{series}{ICSE
  '18})}. \bibinfo{publisher}{Association for Computing Machinery},
  \bibinfo{address}{New York, NY, USA}, \bibinfo{pages}{946–957}.
\newblock
\showISBNx{9781450356381}
\urldef\tempurl%
\url{https://doi.org/10.1145/3180155.3180187}
\showDOI{\tempurl}


\bibitem[\protect\citeauthoryear{Krugler}{Krugler}{2013}]%
        {Krugler2013}
\bibfield{author}{\bibinfo{person}{Ken Krugler}.}
  \bibinfo{year}{2013}\natexlab{}.
\newblock \showarticletitle{Krugle Code Search Architecture}. In
  \bibinfo{booktitle}{\emph{Finding Source Code on the Web for Remix and
  Reuse}}.
\newblock


\bibitem[\protect\citeauthoryear{Le, Nguyen, Le, Phung, Montague, de~Vel, and
  Qu}{Le et~al\mbox{.}}{2019a}]%
        {le2019}
\bibfield{author}{\bibinfo{person}{Tue Le}, \bibinfo{person}{Tuan Nguyen},
  \bibinfo{person}{Trung Le}, \bibinfo{person}{Dinh~Q. Phung},
  \bibinfo{person}{Paul Montague}, \bibinfo{person}{Olivier~Y. de Vel}, {and}
  \bibinfo{person}{Lizhen Qu}.} \bibinfo{year}{2019}\natexlab{a}.
\newblock \showarticletitle{Maximal Divergence Sequential Autoencoder for
  Binary Software Vulnerability Detection}. In \bibinfo{booktitle}{\emph{7th
  International Conference on Learning Representations, {ICLR} 2019, New
  Orleans, LA, USA, May 6-9, 2019}}. \bibinfo{publisher}{OpenReview.net}.
\newblock
\urldef\tempurl%
\url{https://openreview.net/forum?id=ByloIiCqYQ}
\showURL{%
\tempurl}


\bibitem[\protect\citeauthoryear{Le, Nguyen, Le, Phung, Montague, de~Vel, and
  Qu}{Le et~al\mbox{.}}{2019b}]%
        {Le19}
\bibfield{author}{\bibinfo{person}{Tue Le}, \bibinfo{person}{Tuan Nguyen},
  \bibinfo{person}{Trung Le}, \bibinfo{person}{Dinh~Q. Phung},
  \bibinfo{person}{Paul Montague}, \bibinfo{person}{Olivier~Y. de Vel}, {and}
  \bibinfo{person}{Lizhen Qu}.} \bibinfo{year}{2019}\natexlab{b}.
\newblock \showarticletitle{Maximal Divergence Sequential Autoencoder for
  Binary Software Vulnerability Detection}. In \bibinfo{booktitle}{\emph{7th
  International Conference on Learning Representations, {ICLR} 2019, New
  Orleans, LA, USA, May 6-9, 2019}}. \bibinfo{publisher}{OpenReview.net}.
\newblock
\urldef\tempurl%
\url{https://openreview.net/forum?id=ByloIiCqYQ}
\showURL{%
\tempurl}


\bibitem[\protect\citeauthoryear{Lemos, Bajracharya, Ossher, Morla, Masiero,
  Baldi, and Lopes}{Lemos et~al\mbox{.}}{2007}]%
        {Lemos07}
\bibfield{author}{\bibinfo{person}{Ot\'{a}vio Augusto~Lazzarini Lemos},
  \bibinfo{person}{Sushil~Krishna Bajracharya}, \bibinfo{person}{Joel Ossher},
  \bibinfo{person}{Ricardo~Santos Morla}, \bibinfo{person}{Paulo~Cesar
  Masiero}, \bibinfo{person}{Pierre Baldi}, {and}
  \bibinfo{person}{Cristina~Videira Lopes}.} \bibinfo{year}{2007}\natexlab{}.
\newblock \showarticletitle{CodeGenie: Using Test-Cases to Search and Reuse
  Source Code}. In \bibinfo{booktitle}{\emph{Proceedings of the Twenty-Second
  IEEE/ACM International Conference on Automated Software Engineering}}
  (Atlanta, Georgia, USA) \emph{(\bibinfo{series}{ASE '07})}.
  \bibinfo{publisher}{Association for Computing Machinery},
  \bibinfo{address}{New York, NY, USA}, \bibinfo{pages}{525–526}.
\newblock
\showISBNx{9781595938824}
\urldef\tempurl%
\url{https://doi.org/10.1145/1321631.1321726}
\showDOI{\tempurl}


\bibitem[\protect\citeauthoryear{Li, Kim, and Chandra}{Li
  et~al\mbox{.}}{2019a}]%
        {li2019neural}
\bibfield{author}{\bibinfo{person}{Hongyu Li}, \bibinfo{person}{Seohyun Kim},
  {and} \bibinfo{person}{Satish Chandra}.} \bibinfo{year}{2019}\natexlab{a}.
\newblock \bibinfo{title}{Neural Code Search Evaluation Dataset}.
\newblock
\newblock
\showeprint[arxiv]{1908.09804}~[cs.SE]


\bibitem[\protect\citeauthoryear{Li, Kim, and Chandra}{Li
  et~al\mbox{.}}{2019b}]%
        {li2019}
\bibfield{author}{\bibinfo{person}{Hongyu Li}, \bibinfo{person}{Seohyun Kim},
  {and} \bibinfo{person}{Satish Chandra}.} \bibinfo{year}{2019}\natexlab{b}.
\newblock \showarticletitle{Neural Code Search Evaluation Dataset}.
\newblock \bibinfo{journal}{\emph{CoRR}}  \bibinfo{volume}{abs/1908.09804}
  (\bibinfo{year}{2019}).
\newblock


\bibitem[\protect\citeauthoryear{Li, Wang, Lyu, and King}{Li
  et~al\mbox{.}}{2018}]%
        {li2018}
\bibfield{author}{\bibinfo{person}{Jian Li}, \bibinfo{person}{Yue Wang},
  \bibinfo{person}{Michael~R. Lyu}, {and} \bibinfo{person}{Irwin King}.}
  \bibinfo{year}{2018}\natexlab{}.
\newblock \showarticletitle{Code Completion with Neural Attention and Pointer
  Networks}. In \bibinfo{booktitle}{\emph{Proceedings of the Twenty-Seventh
  International Joint Conference on Artificial Intelligence, {IJCAI-18}}}.
  \bibinfo{publisher}{International Joint Conferences on Artificial
  Intelligence Organization}, \bibinfo{pages}{4159--4165}.
\newblock
\urldef\tempurl%
\url{https://doi.org/10.24963/ijcai.2018/578}
\showDOI{\tempurl}


\bibitem[\protect\citeauthoryear{Li and K\"{o}nig}{Li and K\"{o}nig}{2011}]%
        {Ping11}
\bibfield{author}{\bibinfo{person}{Ping Li} {and}
  \bibinfo{person}{Arnd~Christian K\"{o}nig}.} \bibinfo{year}{2011}\natexlab{}.
\newblock \showarticletitle{Theory and Applications of b-Bit Minwise Hashing}.
\newblock \bibinfo{journal}{\emph{Commun. ACM}} \bibinfo{volume}{54},
  \bibinfo{number}{8} (\bibinfo{date}{Aug.} \bibinfo{year}{2011}),
  \bibinfo{pages}{101–109}.
\newblock
\showISSN{0001-0782}
\urldef\tempurl%
\url{https://doi.org/10.1145/1978542.1978566}
\showDOI{\tempurl}


\bibitem[\protect\citeauthoryear{Li and König}{Li and König}{2010}]%
        {Li10}
\bibfield{author}{\bibinfo{person}{Ping Li} {and}
  \bibinfo{person}{Arnd~Christian König}.} \bibinfo{year}{2010}\natexlab{}.
\newblock \showarticletitle{b-Bit Minwise Hashing}. In
  \bibinfo{booktitle}{\emph{Nineteenth International World Wide Web Conference
  (WWW 2010)} (\bibinfo{edition}{nineteenth international world wide web
  conference (www 2010)} ed.)}.
\newblock


\bibitem[\protect\citeauthoryear{Li, Qin, Yan, Shen, and Chen}{Li
  et~al\mbox{.}}{2020}]%
        {Wei20}
\bibfield{author}{\bibinfo{person}{Wei Li}, \bibinfo{person}{Haozhe Qin},
  \bibinfo{person}{Shuhan Yan}, \bibinfo{person}{Beijun Shen}, {and}
  \bibinfo{person}{Yuting Chen}.} \bibinfo{year}{2020}\natexlab{}.
\newblock \showarticletitle{Learning Code-Query Interaction for Enhancing Code
  Searches}. In \bibinfo{booktitle}{\emph{2020 IEEE International Conference on
  Software Maintenance and Evolution (ICSME)}}. \bibinfo{pages}{115--126}.
\newblock
\urldef\tempurl%
\url{https://doi.org/10.1109/ICSME46990.2020.00021}
\showDOI{\tempurl}


\bibitem[\protect\citeauthoryear{Liu, Xia, Lo, Gao, Yang, and Grundy}{Liu
  et~al\mbox{.}}{2020}]%
        {liu2020opportunities}
\bibfield{author}{\bibinfo{person}{Chao Liu}, \bibinfo{person}{Xin Xia},
  \bibinfo{person}{David Lo}, \bibinfo{person}{Cuiyun Gao},
  \bibinfo{person}{Xiaohu Yang}, {and} \bibinfo{person}{John Grundy}.}
  \bibinfo{year}{2020}\natexlab{}.
\newblock \bibinfo{title}{Opportunities and Challenges in Code Search Tools}.
\newblock
\newblock
\showeprint[arxiv]{2011.02297}~[cs.SE]


\bibitem[\protect\citeauthoryear{Luan, Yang, Barnaby, Sen, and Chandra}{Luan
  et~al\mbox{.}}{2019}]%
        {luan2019}
\bibfield{author}{\bibinfo{person}{Sifei Luan}, \bibinfo{person}{Di Yang},
  \bibinfo{person}{Celeste Barnaby}, \bibinfo{person}{Koushik Sen}, {and}
  \bibinfo{person}{Satish Chandra}.} \bibinfo{year}{2019}\natexlab{}.
\newblock \showarticletitle{Aroma: Code Recommendation via Structural Code
  Search}.
\newblock \bibinfo{journal}{\emph{Proc. ACM Program. Lang.}}
  \bibinfo{volume}{3}, \bibinfo{number}{OOPSLA}, Article
  \bibinfo{articleno}{152} (\bibinfo{date}{Oct.} \bibinfo{year}{2019}),
  \bibinfo{numpages}{28}~pages.
\newblock


\bibitem[\protect\citeauthoryear{Ondov, Treangen, Melsted, Mallonee, Bergman,
  Koren, and Phillippy}{Ondov et~al\mbox{.}}{2016}]%
        {Ondov16}
\bibfield{author}{\bibinfo{person}{Brian Ondov}, \bibinfo{person}{Todd
  Treangen}, \bibinfo{person}{Páll Melsted}, \bibinfo{person}{Adam Mallonee},
  \bibinfo{person}{Nicholas Bergman}, \bibinfo{person}{Sergey Koren}, {and}
  \bibinfo{person}{Adam Phillippy}.} \bibinfo{year}{2016}\natexlab{}.
\newblock \showarticletitle{Mash: Fast genome and metagenome distance
  estimation using MinHash}.
\newblock \bibinfo{journal}{\emph{Genome Biology}}  \bibinfo{volume}{17}
  (\bibinfo{date}{06} \bibinfo{year}{2016}).
\newblock
\urldef\tempurl%
\url{https://doi.org/10.1186/s13059-016-0997-x}
\showDOI{\tempurl}


\bibitem[\protect\citeauthoryear{Pacheco, Lahiri, Ernst, and Ball}{Pacheco
  et~al\mbox{.}}{2007}]%
        {PachecoLEB2007}
\bibfield{author}{\bibinfo{person}{Carlos Pacheco},
  \bibinfo{person}{Shuvendu~K. Lahiri}, \bibinfo{person}{Michael~D. Ernst},
  {and} \bibinfo{person}{Thomas Ball}.} \bibinfo{year}{2007}\natexlab{}.
\newblock \showarticletitle{Feedback-directed random test generation}. In
  \bibinfo{booktitle}{\emph{ICSE 2007, Proceedings of the 29th International
  Conference on Software Engineering}}. \bibinfo{address}{Minneapolis, MN,
  USA}, \bibinfo{pages}{75--84}.
\newblock


\bibitem[\protect\citeauthoryear{Petrovic}{Petrovic}{2013}]%
        {Petrovic2013RealtimeED}
\bibfield{author}{\bibinfo{person}{S. Petrovic}.}
  \bibinfo{year}{2013}\natexlab{}.
\newblock \showarticletitle{Real-time event detection in massive streams}.
\newblock


\bibitem[\protect\citeauthoryear{Pradel and Sen}{Pradel and Sen}{2018}]%
        {Pradel2018}
\bibfield{author}{\bibinfo{person}{Michael Pradel} {and}
  \bibinfo{person}{Koushik Sen}.} \bibinfo{year}{2018}\natexlab{}.
\newblock \showarticletitle{DeepBugs: A Learning Approach to Name-Based Bug
  Detection}.
\newblock \bibinfo{journal}{\emph{Proc. ACM Program. Lang.}}
  \bibinfo{volume}{2}, \bibinfo{number}{OOPSLA}, Article
  \bibinfo{articleno}{147} (\bibinfo{date}{Oct.} \bibinfo{year}{2018}),
  \bibinfo{numpages}{25}~pages.
\newblock
\urldef\tempurl%
\url{https://doi.org/10.1145/3276517}
\showDOI{\tempurl}


\bibitem[\protect\citeauthoryear{Puri, Kung, Janssen, Zhang, Domeniconi,
  Zolotov, Dolby, Chen, Choudhury, Decker, Thost, Buratti, Pujar, and
  Finkler}{Puri et~al\mbox{.}}{2021}]%
        {puri2021codenet}
\bibfield{author}{\bibinfo{person}{Ruchir Puri}, \bibinfo{person}{David Kung},
  \bibinfo{person}{Geert Janssen}, \bibinfo{person}{Wei Zhang},
  \bibinfo{person}{Giacomo Domeniconi}, \bibinfo{person}{Vladmir Zolotov},
  \bibinfo{person}{Julian Dolby}, \bibinfo{person}{Jie Chen},
  \bibinfo{person}{Mihir Choudhury}, \bibinfo{person}{Lindsey Decker},
  \bibinfo{person}{Veronika Thost}, \bibinfo{person}{Luca Buratti},
  \bibinfo{person}{Saurabh Pujar}, {and} \bibinfo{person}{Ulrich Finkler}.}
  \bibinfo{year}{2021}\natexlab{}.
\newblock \showarticletitle{Project CodeNet: A Large-Scale AI for Code Dataset
  for Learning a Diversity of Coding Tasks}.
\newblock


\bibitem[\protect\citeauthoryear{Rajaraman and Ullman}{Rajaraman and
  Ullman}{2012}]%
        {rajaraman2012mining}
\bibfield{author}{\bibinfo{person}{Anand Rajaraman} {and}
  \bibinfo{person}{Jeffrey~David Ullman}.} \bibinfo{year}{2012}\natexlab{}.
\newblock \bibinfo{booktitle}{\emph{Mining of massive datasets}}.
\newblock \bibinfo{publisher}{Cambridge University Press},
  \bibinfo{address}{Cambridge}.
\newblock
\showISBNx{9781139157926 1139157922 9781107015357 1107015359}
\urldef\tempurl%
\url{http://www.amazon.de/Mining-Massive-Datasets-Anand-Rajaraman/dp/1107015359/ref=sr_1_1?ie=UTF8&qid=1350890245&sr=8-1}
\showURL{%
\tempurl}


\bibitem[\protect\citeauthoryear{Rangwala and Rasheed}{Rangwala and
  Rasheed}{2013}]%
        {Rangwala2013MCMinHMC}
\bibfield{author}{\bibinfo{person}{Huzefa Rangwala} {and}
  \bibinfo{person}{Zeehasham Rasheed}.} \bibinfo{year}{2013}\natexlab{}.
\newblock \showarticletitle{MC-MinH: Metagenome Clustering using Minwise based
  Hashing}. In \bibinfo{booktitle}{\emph{SDM}}.
\newblock


\bibitem[\protect\citeauthoryear{Rong, Yoon, Bergen, Elezabi, Bailis, Levis,
  and Beroza}{Rong et~al\mbox{.}}{2018}]%
        {Rong18}
\bibfield{author}{\bibinfo{person}{Kexin Rong}, \bibinfo{person}{Clara~E.
  Yoon}, \bibinfo{person}{Karianne~J. Bergen}, \bibinfo{person}{Hashem
  Elezabi}, \bibinfo{person}{Peter Bailis}, \bibinfo{person}{Philip Levis},
  {and} \bibinfo{person}{Gregory~C. Beroza}.} \bibinfo{year}{2018}\natexlab{}.
\newblock \showarticletitle{Locality-Sensitive Hashing for Earthquake
  Detection: A Case Study of Scaling Data-Driven Science}.
\newblock \bibinfo{journal}{\emph{Proc. VLDB Endow.}} \bibinfo{volume}{11},
  \bibinfo{number}{11} (\bibinfo{date}{July} \bibinfo{year}{2018}),
  \bibinfo{pages}{1674–1687}.
\newblock
\showISSN{2150-8097}


\bibitem[\protect\citeauthoryear{Saini, Farmahinifarahani, Lu, Baldi, and
  Lopes}{Saini et~al\mbox{.}}{2018}]%
        {Saini18}
\bibfield{author}{\bibinfo{person}{Vaibhav Saini}, \bibinfo{person}{Farima
  Farmahinifarahani}, \bibinfo{person}{Yadong Lu}, \bibinfo{person}{Pierre
  Baldi}, {and} \bibinfo{person}{Cristina~V. Lopes}.}
  \bibinfo{year}{2018}\natexlab{}.
\newblock \showarticletitle{Oreo: Detection of Clones in the Twilight Zone}. In
  \bibinfo{booktitle}{\emph{Proceedings of the 2018 26th ACM Joint Meeting on
  European Software Engineering Conference and Symposium on the Foundations of
  Software Engineering}} (Lake Buena Vista, FL, USA)
  \emph{(\bibinfo{series}{ESEC/FSE 2018})}. \bibinfo{publisher}{Association for
  Computing Machinery}, \bibinfo{address}{New York, NY, USA},
  \bibinfo{pages}{354–365}.
\newblock
\showISBNx{9781450355735}


\bibitem[\protect\citeauthoryear{{Sajnani}, {Saini}, {Svajlenko}, {Roy}, and
  {Lopes}}{{Sajnani} et~al\mbox{.}}{2016}]%
        {Sajani16}
\bibfield{author}{\bibinfo{person}{H. {Sajnani}}, \bibinfo{person}{V. {Saini}},
  \bibinfo{person}{J. {Svajlenko}}, \bibinfo{person}{C.~K. {Roy}}, {and}
  \bibinfo{person}{C.~V. {Lopes}}.} \bibinfo{year}{2016}\natexlab{}.
\newblock \showarticletitle{SourcererCC: Scaling Code Clone Detection to
  Big-Code}. In \bibinfo{booktitle}{\emph{2016 IEEE/ACM 38th International
  Conference on Software Engineering (ICSE)}}. \bibinfo{pages}{1157--1168}.
\newblock
\urldef\tempurl%
\url{https://doi.org/10.1145/2884781.2884877}
\showDOI{\tempurl}


\bibitem[\protect\citeauthoryear{Shrivastava and Li}{Shrivastava and
  Li}{2015}]%
        {shrivastava2015asymmetric}
\bibfield{author}{\bibinfo{person}{Anshumali Shrivastava} {and}
  \bibinfo{person}{Ping Li}.} \bibinfo{year}{2015}\natexlab{}.
\newblock \showarticletitle{Asymmetric minwise hashing for indexing binary
  inner products and set containment}. In \bibinfo{booktitle}{\emph{Proceedings
  of the 24th international conference on world wide web}}.
  \bibinfo{pages}{981--991}.
\newblock


\bibitem[\protect\citeauthoryear{Tufano, Drain, Svyatkovskiy, Deng, and
  Sundaresan}{Tufano et~al\mbox{.}}{2020}]%
        {tufano2020unit}
\bibfield{author}{\bibinfo{person}{Michele Tufano}, \bibinfo{person}{Dawn
  Drain}, \bibinfo{person}{Alexey Svyatkovskiy}, \bibinfo{person}{Shao~Kun
  Deng}, {and} \bibinfo{person}{Neel Sundaresan}.}
  \bibinfo{year}{2020}\natexlab{}.
\newblock \bibinfo{title}{Unit Test Case Generation with Transformers}.
\newblock
\newblock
\showeprint[arxiv]{2009.05617}~[cs.SE]


\bibitem[\protect\citeauthoryear{{Ueda}, {Kamiya}, {Kusumoto}, and
  {Inoue}}{{Ueda} et~al\mbox{.}}{2002}]%
        {Ueda02}
\bibfield{author}{\bibinfo{person}{Y. {Ueda}}, \bibinfo{person}{T. {Kamiya}},
  \bibinfo{person}{S. {Kusumoto}}, {and} \bibinfo{person}{K. {Inoue}}.}
  \bibinfo{year}{2002}\natexlab{}.
\newblock \showarticletitle{On detection of gapped code clones using gap
  locations}. In \bibinfo{booktitle}{\emph{Ninth Asia-Pacific Software
  Engineering Conference, 2002.}} \bibinfo{pages}{327--336}.
\newblock
\urldef\tempurl%
\url{https://doi.org/10.1109/APSEC.2002.1183002}
\showDOI{\tempurl}


\bibitem[\protect\citeauthoryear{{White}, {Tufano}, {Vendome}, and
  {Poshyvanyk}}{{White} et~al\mbox{.}}{2016}]%
        {White16}
\bibfield{author}{\bibinfo{person}{M. {White}}, \bibinfo{person}{M. {Tufano}},
  \bibinfo{person}{C. {Vendome}}, {and} \bibinfo{person}{D. {Poshyvanyk}}.}
  \bibinfo{year}{2016}\natexlab{}.
\newblock \showarticletitle{Deep learning code fragments for code clone
  detection}. In \bibinfo{booktitle}{\emph{2016 31st IEEE/ACM International
  Conference on Automated Software Engineering (ASE)}}.
  \bibinfo{pages}{87--98}.
\newblock


\bibitem[\protect\citeauthoryear{Yan, Yu, Chen, Shen, and Jiang}{Yan
  et~al\mbox{.}}{2020}]%
        {Shuhan20}
\bibfield{author}{\bibinfo{person}{Shuhan Yan}, \bibinfo{person}{Hang Yu},
  \bibinfo{person}{Yuting Chen}, \bibinfo{person}{Beijun Shen}, {and}
  \bibinfo{person}{Lingxiao Jiang}.} \bibinfo{year}{2020}\natexlab{}.
\newblock \showarticletitle{Are the Code Snippets What We Are Searching for? A
  Benchmark and an Empirical Study on Code Search with Natural-Language
  Queries}. In \bibinfo{booktitle}{\emph{2020 IEEE 27th International
  Conference on Software Analysis, Evolution and Reengineering (SANER)}}.
  \bibinfo{pages}{344--354}.
\newblock
\urldef\tempurl%
\url{https://doi.org/10.1109/SANER48275.2020.9054840}
\showDOI{\tempurl}


\bibitem[\protect\citeauthoryear{Zamani and Croft}{Zamani and Croft}{2018}]%
        {Zamani18}
\bibfield{author}{\bibinfo{person}{Hamed Zamani} {and}
  \bibinfo{person}{W.~Bruce Croft}.} \bibinfo{year}{2018}\natexlab{}.
\newblock \showarticletitle{On the Theory of Weak Supervision for Information
  Retrieval}. In \bibinfo{booktitle}{\emph{Proceedings of the 2018 ACM SIGIR
  International Conference on Theory of Information Retrieval}} (Tianjin,
  China) \emph{(\bibinfo{series}{ICTIR '18})}. \bibinfo{publisher}{Association
  for Computing Machinery}, \bibinfo{address}{New York, NY, USA},
  \bibinfo{pages}{147–154}.
\newblock
\showISBNx{9781450356565}
\urldef\tempurl%
\url{https://doi.org/10.1145/3234944.3234968}
\showDOI{\tempurl}


\bibitem[\protect\citeauthoryear{Zhou and Sharma}{Zhou and Sharma}{2017}]%
        {Zhou17}
\bibfield{author}{\bibinfo{person}{Yaqin Zhou} {and} \bibinfo{person}{Asankhaya
  Sharma}.} \bibinfo{year}{2017}\natexlab{}.
\newblock \showarticletitle{Automated Identification of Security Issues from
  Commit Messages and Bug Reports}. In \bibinfo{booktitle}{\emph{Proceedings of
  the 2017 11th Joint Meeting on Foundations of Software Engineering}}
  (Paderborn, Germany) \emph{(\bibinfo{series}{ESEC/FSE 2017})}.
  \bibinfo{publisher}{Association for Computing Machinery},
  \bibinfo{address}{New York, NY, USA}, \bibinfo{pages}{914–919}.
\newblock
\showISBNx{9781450351058}
\urldef\tempurl%
\url{https://doi.org/10.1145/3106237.3117771}
\showDOI{\tempurl}


\bibitem[\protect\citeauthoryear{Zhu, Nargesian, Pu, and Miller}{Zhu
  et~al\mbox{.}}{2016}]%
        {Zhu16}
\bibfield{author}{\bibinfo{person}{Erkang Zhu}, \bibinfo{person}{Fatemeh
  Nargesian}, \bibinfo{person}{Ken~Q. Pu}, {and} \bibinfo{person}{Ren\'{e}e~J.
  Miller}.} \bibinfo{year}{2016}\natexlab{}.
\newblock \showarticletitle{LSH Ensemble: Internet-Scale Domain Search}.
\newblock \bibinfo{journal}{\emph{Proc. VLDB Endow.}} \bibinfo{volume}{9},
  \bibinfo{number}{12} (\bibinfo{date}{Aug.} \bibinfo{year}{2016}),
  \bibinfo{pages}{1185–1196}.
\newblock
\showISSN{2150-8097}


\bibitem[\protect\citeauthoryear{Zobel}{Zobel}{1998}]%
        {Zobel98}
\bibfield{author}{\bibinfo{person}{Justin Zobel}.}
  \bibinfo{year}{1998}\natexlab{}.
\newblock \showarticletitle{How Reliable Are the Results of Large-Scale
  Information Retrieval Experiments?}. In \bibinfo{booktitle}{\emph{Proceedings
  of the 21st Annual International ACM SIGIR Conference on Research and
  Development in Information Retrieval}} (Melbourne, Australia)
  \emph{(\bibinfo{series}{SIGIR '98})}. \bibinfo{publisher}{Association for
  Computing Machinery}, \bibinfo{address}{New York, NY, USA},
  \bibinfo{pages}{307–314}.
\newblock
\showISBNx{1581130155}


\bibitem[\protect\citeauthoryear{Z{\"u}gner, Kirschstein, Catasta, Leskovec,
  and G{\"u}nnemann}{Z{\"u}gner et~al\mbox{.}}{2021}]%
        {zuegner_code_transformer_2021}
\bibfield{author}{\bibinfo{person}{Daniel Z{\"u}gner}, \bibinfo{person}{Tobias
  Kirschstein}, \bibinfo{person}{Michele Catasta}, \bibinfo{person}{Jure
  Leskovec}, {and} \bibinfo{person}{Stephan G{\"u}nnemann}.}
  \bibinfo{year}{2021}\natexlab{}.
\newblock \showarticletitle{Language-Agnostic Representation Learning of Source
  Code from Structure and Context}. In \bibinfo{booktitle}{\emph{International
  Conference on Learning Representations (ICLR)}}.
\newblock


\end{thebibliography}

\appendix

\end{document}